\documentclass[journal]{IEEEtai}

\usepackage[colorlinks,urlcolor=blue,linkcolor=blue,citecolor=blue]{hyperref}

\usepackage{color,array}

\usepackage{algorithmic}
\usepackage{textcomp}
\usepackage{microtype}
\usepackage{graphicx}
\usepackage{subfigure}
\usepackage{tabularx} 
\usepackage[table]{xcolor} 
\usepackage{booktabs} 
\usepackage{algorithm}
\usepackage{amsmath}
\usepackage[colorlinks,urlcolor=blue,linkcolor=blue,citecolor=blue]{hyperref}

\usepackage{amsmath}
\usepackage{amssymb}
\usepackage{mathtools}
\usepackage{amsthm}
\usepackage[numbers]{natbib}

\begin{document}

\title{ GetNetUPAM: Ecologically Informed Nested Cross‑Validation and Noise‑Robust Attention for Marine Bioacoustic Monitoring }

\author{Nicholas R. Rasmussen, Rodrigue Rizk, Longwei Wang, KC Santosh}

\maketitle

\begin{abstract}

Deploying reliable bioacoustic monitoring systems requires models that generalize under high‑noise, low-SNR conditions and evaluation protocols that expose deployment‑relevant failure modes—gaps largely unaddressed in current UPAM practice. Intrinsic noise, variable propagation, and mixed biological and anthropogenic sources induce distribution shifts that conventional models and single‑split evaluations obscure, inflating performance and masking instability. We introduce \textit{GetNetUPAM}, a hierarchical nested cross‑validation framework that uses the nested stage to \textbf{quantify model stability} rather than tune for inflated hold‑out scores. By partitioning data into site--year blocks, GetNetUPAM preserves ecological heterogeneity and forces each outer fold to represent a distinct environmental regime, preventing overfitting to localized noise or sensor artifacts. Inner stratified folds measure generalization across the full UPAM signal distribution, enforcing strict separation between model development and the outer held‑out deployment condition. Using GetNetUPAM, we evaluate the \textit{Adaptive Resolution Pooling and Attention Network (ARPA‑N)}, a CNN architecture for irregular spectrogram dimensions. ARPA‑N integrates CBAM spatial attention as a learned noise suppressor, producing attention maps that localize true call structure and avoid the global, non‑biological cues exploited by standard CNNs on long‑window data. Under GetNetUPAM, ARPA‑N generalizes robustly across diverse environmental regimes. In the zero training support Balleny Islands region, it reduces false positives per hour by over an order-of-magnitude ($\sim$10$\times$) at fixed 90\% recall, yielding consistently improved metrics across folds. These advances provide a reproducible benchmark and move UPAM toward scalable, deployment‑reliable ecological monitoring.

\end{abstract}

\begin{IEEEImpStatement}

GetNetUPAM supports non-invasive monitoring of marine species such as Blue Whales, enabling conservation efforts while minimizing disturbance and promoting ethical research. Stability under noisy underwater conditions reduces manual annotation effort and improves detector reliability. By modeling environmental variability within a hierarchical evaluation, GetNetUPAM provides a clearer picture of system behavior across differing site--year conditions without assuming broad geographic generalization. The ARPA-N architecture strengthens performance by normalizing inputs through adaptive pooling and using CBAM spatial attention as a learned noise suppressor, helping the model emphasize biologically meaningful call structure in challenging acoustic environments.

\end{IEEEImpStatement}


\section{Introduction}

\begin{figure}[t!]
\centering
\includegraphics[width=.95\linewidth]{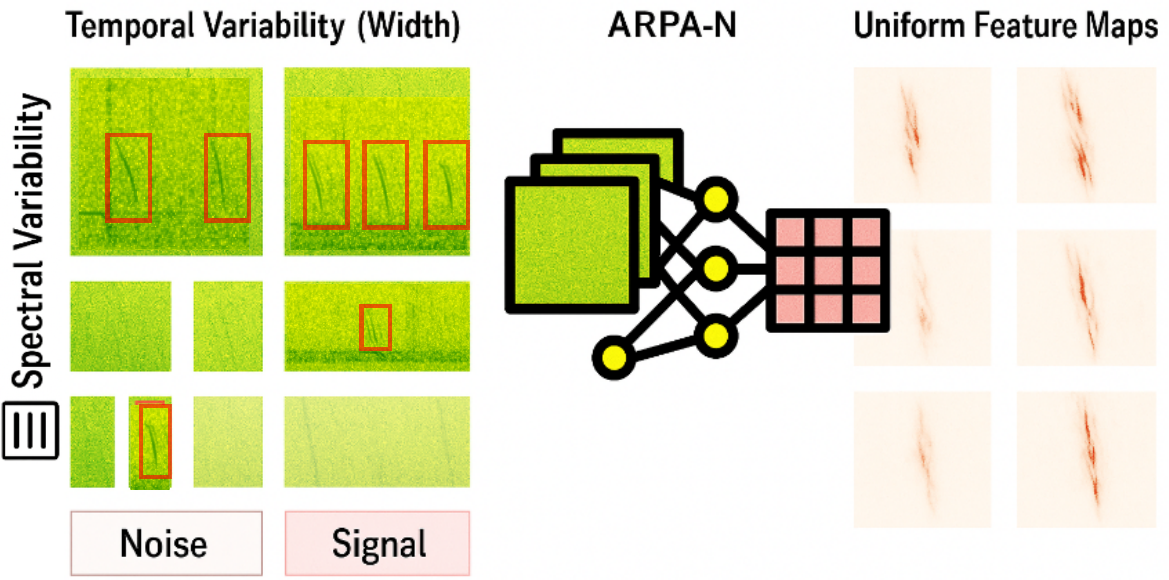}
\caption{
\textbf{Motivation for GetNetUPAM and ARPA-N.}
(\emph{Left}) UPAM spectrograms vary widely in resolution and aspect ratio, with whale calls (\textcolor{red}{red}) embedded in heterogeneous noise (\textcolor{green}{green}), complicating stability and masking performance variability. (\emph{Center}) ARPA‑N uses CBAM spatial attention~\cite{woo2018cbam} to suppress noise and normalize intermediate representations. (\emph{Right}) The resulting uniform feature maps preserve call structure while reducing cross‑condition variability, enabling rigorous UPAM evaluation for conservation~\cite{Parsons2022}.}
\label{fig:gF}
\end{figure}

Deploying reliable UPAM systems is challenging due to strong spatiotemporal variability, shifting noise floors, and mixed biological and anthropogenic sources, all of which create distributional heterogeneity that static train--test splits fail to capture \cite{geirhos2020shortcut, varma2006bias, Miller2021, Tuia2021, Minello2021, Valavi2019}. Climate‑driven ecosystem change further increases the need for stable acoustic monitoring, particularly for species whose distributions are primarily inferred from sound \cite{Foden2013, Hare2016, Hildebrand2024, Balcazar2015}.

Progress is limited by two gaps. The evaluation gap arises because random‑subset benchmarks conflate memorization of site‑specific noise with genuine robustness and provide no fold‑level variance estimates \cite{Roberts2017, Miller2022}. Documented failures under site shifts \cite{Haver2023} and incomplete reporting in site‑blocked studies \cite{Schall2024} leave stability unresolved. On the other hand, the architecture gap stems from UPAM pipelines that produce irregular, variable‑aspect spectrograms \cite{Miller2022, Rasmussen2024}, which standard CNNs, generally designed for fixed input geometry, model inefficiently with susceptibility to shortcut cues \cite{geirhos2020shortcut}.

\begin{table*}[t!]
\centering
\resizebox{\linewidth}{!}{%
\begin{tabular}{llllll}
\hline \rowcolor{gray!15}
\textbf{Method} & \textbf{Call Type} & \textbf{Sample Window} & \textbf{Key Metric} & \textbf{Result} & \textbf{Comparability} \\
\hline
Schall et al.\ \cite{Schall2024} & Blue + Fin & 15 s & Overall Fitness & .830 & Partial (no AP/F1/stability) \\
Nihal et al.\ \cite{nihal2025} & All (binary) & 60--1800\,s & Binary F1 / Loc.\ Prec. & 0.81 / 0.64 & Partial (binary, no D-call / stability) \\
WhaleVADBPN \cite{whalevad2025} & D-Call & Strong-label subset & Event-level F1 & .22--.34 & Partial (poor performance, no stability) \\
Miller et al.\ \cite{Miller2022} & D-Call & 4.5 s & R / P / F1 & .758 / .658 / .704  & Partial (no stability, no cross-site) \\
Rasmussen et al.\ \cite{Rasmussen2024} & Blue + Fin + D-Call & Variable & R / P / AP & .991 / .684 / .817 & Partial (no stability, no cross-site) \\
Babalola et al. \cite{babalola2024wavelet} & Anthem-calls & 4 s & Event-level F1 & .86\% & Limited (single-site, no cross-site) \\
Jancovich et al.\ \cite{jancovich2026} & Z-call (only) & 60 s & R / P / F1 & .87 / .65 / .75 & Limited (single-site, high-SNR) \\
Baumgarte et al. \cite{baumgarte2024fin} & Fin 20\,Hz & 5\,s & R / P / FP/hr & .83 / .918 / 12.1 & Limited (single-site, no cross-site) \\
Rademan et al. \cite{waveletscattering2024} & A and D Calls\,Hz & 5\,s & P/R Curves & Visual & Limited (single-site, no cross-site) \\
Parry et al.\ \cite{parry2025} & Fin whale & 3\,s & Cross-env.\ gain & Relative only & Limited (no ATBFL quantitative test) \\
MONSTER \cite{monster2025} & All 7 types & 10 s & Loss & Obscured & Limited (fixed split, no stability) \\
BioDCASE \cite{biodcase2025} & All 7 types & 11.8\,s & R / P / F1 & .477 / .525 / .500 & Limited (fixed split, no stability) \\
Van Toor \& Vaz \cite{Toor2025} & ABZ calls & 11.8\,s & R / P & .28 / .70 & Limited (BioDCASE split, no stability) \\

\hline
\end{tabular}%
}
\caption{Comparison of recent ATBFL quantitative results. Comparability reflects evaluation compatibility with GetNetUPAM.}
\label{tab:RW_comparison}
\end{table*}

To address the evaluation gap, we introduce \textit{GetNetUPAM}, the first UPAM framework to use nested cross‑validation for stability quantification rather than performance inflation \cite{varma2006bias}. Unlike conventional site‑block CV, where repeated inspection of the held‑out block implicitly adapts models, thresholds, and hyperparameters to that specific block, GetNetUPAM enforces strict separation between development and evaluation. Site‑year blocking mirrors ecological and deployment‑like variability, while the inner loop characterizes precision–stability tradeoffs across noise and SNR regimes without using the outer block for model selection. This yields interpretable robustness estimates that scalar metrics obscure, reducing optimistic bias from reusing the same block throughout model iteration.

To address the architecture gap, we propose \textit{ARPA‑N}, which combines adaptive resolution pooling, standardizing irregular spectrogram dimensions while expanding receptive fields, with CBAM spatial attention as a learned noise suppressor. Adaptive pooling stabilizes long‑range structure without transformer‑level cost, while attention reduces reliance on global, non‑biological artifacts \cite{geirhos2020shortcut}. Under GetNetUPAM, ARPA‑N improves AP by 14.7\% and reduces false positives per hour (FP/hr) at 90\% Recall by an order-of-magnitude relative to 60s baselines while also improving stability, directly improving metrics relevant to conservation decision‑making \cite{Schall2024}.

Our contributions are summarized as follows:
\begin{itemize}
    \item \textbf{GetNetUPAM} combines nested and site‑blocked CV to measure stability and close the evaluation gap ensuring deployment readiness across different sites.
    \item \textbf{ARPA‑N} handles resolution heterogeneity via adaptive pooling and spatial attention, improving performance and robustness without resampling.
    \item  Embeddings show that CBAM spatial attention suppresses non‑target global cues, mitigating shortcut learning and improving robustness, a first in ecology.
    \item  ARPA‑N’s modular design supports full‑depth and edge‑class variants (e.g., All$-$D) for constraint‑driven deployments where efficiency is critical.
\end{itemize}

\section{Related Work} \label{rel}

The Antarctic Blue and Fin Whale Acoustic Trends Project (ATBFL) \cite{Miller2021} is the most comprehensive circumpolar repository of baleen whale recordings, with 1{,}880.25 annotated hours and over 300{,}000 hours of supplemental audio. We use “site-years” that applied systematic random subsampling to extract $\sim$200 hours per site-year with broad diurnal and seasonal coverage. Each site contributes 10–18 annotated hours per month, evenly spanning all 24 hours, while multiple geographically and technically distinct Antarctic sites provide strong cross-site diversity. Strict site-year blocking ensures evaluation on unseen instruments, locations, noise conditions, and seasonal ranges, closely simulating real-world deployment. This scale and heterogeneity underscore core UPAM challenges: detecting sparse, low-frequency, low-SNR blue whale D-calls that resemble other frequency-modulated signals \cite{Miller2021, Miller2022}. ATBFL is now the de facto benchmark for Antarctic baleen detection, formalized in Schall et al.\ \cite{Schall2024} and the BioDCASE challenge \cite{biodcase2025}. Table~\ref{tab:RW_comparison} compares ATBFL methods to GetNetUPAM.

\paragraph{Classical UPAM Approaches} Early pipelines used correlation kernels \cite{Miller2021} or feature‑engineered decision trees comparing spectral energy, SNR, and dynamic changes \cite{Schall2022}. These methods degrade under frequency shifts, seasonal variability, annotator inconsistency, and long 40--50 call features that violate stationarity assumptions \cite{Sirovic2015, Rankin2023, Sirovic2013}. Wavelet- and entropy-based approaches on ATBFL, WT-HMM \cite{babalola2024wavelet} and multiscale entropy with GMMs \cite{babalola2023}, report strong single-site accuracy, but accuracy obscures class imbalance and is not comparable to detection metrics. A wavelet-scattering pipeline \cite{waveletscattering2024} attributes STFT underperformance to frequency averaging, though short 5--9\,s windows likely miss the 15--30$+$\,s structure of blue whale anthem calls \cite{Sirovic2015, Rankin2023, Rasmussen2024, McCallum2018}. Román-Ruiz et al.\ \cite{romanruiz2024} applied circle detection to fin whale pulses at Casey, reporting ${\sim}95$\% recall but omitting precision, F1, and cross-site evaluation. Overall, many classical methods rely on a single-site and accuracy-based reporting, limiting generalization insight.

\paragraph{Early Deep Learning in UPAM} Deep models improved robustness over classical pipelines. Recurrent CNNs classified 9\,s segments \cite{Rasmussen2021}, while DenseNet models on 4.5\,s windows achieved strong hold-out performance and outperformed human annotators \cite{Miller2022}. Rasmussen et al.\ \cite{Rasmussen2024} used a ResNet-18 with varied spectral representations with high recall, though narrow receptive fields limit long-range modeling. Baumgarte \cite{baumgarte2024fin} trained on 5\,s patches for fin whale 20\,Hz pulses using an odd/even-hour split at a single site; strong within-site precision is reported, but lack of cross-site testing limits generalizability. WhaleVADBPN \cite{whalevad2025} refined temporal boundaries on the ATBFL strong-label subset, reporting event-level F1 of 0.22--0.34 under three-fold site-blocked CV.

\paragraph{Transfer Learning and Data-Efficient Detection} Annotation scarcity motivates augmentation and transfer learning. Jancovich et al.\ \cite{jancovich2026} generated semi-synthetic ATBFL training data from a single exemplar and fine-tuned a pretrained speech detector, achieving Recall\,=\,87\%, Precision\,=\,65\%, and F1\,=\,75\% on blue whale Z-calls, though restricted to high-SNR, stereotyped events. Parry et al.\ \cite{parry2025} used a latent diffusion model to blend fin whale calls into Antarctic and Bermudian noise, improving cross-environment generalization over classical mixing, but used 3\,s windows and did not evaluate on ATBFL. Fourie et al.\ \cite{fourie2022neural} applied speech-processing architectures to presegmented ATBFL clips, reporting accuracy only, without continuous-audio evaluation or detection metrics, limiting comparability to real scenarios.

\paragraph{Weakly Supervised and Challenge-Scale Detection}
Nihal et al.\ \cite{nihal2025} introduced a weakly supervised MIL system over the full 1{,}880-hour ATBFL, using 60--1{,}800\,s bags and a dual-stream temporal+spectrogram model trained on presence/absence labels but evaluated with strong labels; their best model reached F1$\approx$0.81 and localization precision $\approx$0.64 under site-year blocked CV, though collapsing all call types into one class limits ecological interpretability. The BioDCASE challenge \cite{biodcase2025} formalized ATBFL as a benchmark, with Task~2 using ATBFL \cite{Miller2021} recordings across seven call types; its ResNet-18 baseline relies on a fixed split without variance estimates or cross-site analysis. Van Toor and Vaz \cite{Toor2025} deployed an ultra-lightweight CNN (8k--120k parameters) on the same split, reporting 70\% precision but only 28\% recall for ABZ calls and 10\% for downsweeps (including D-calls); the 11.8\,s window truncates long calls, and the fixed-split protocol again precludes stability assessment and breaks hierarchical structure.

\paragraph{Cross-Domain Repurposing of ATBFL} 
ATBFL has been repurposed for general time-series classification via MONSTER \cite{monster2025}, which extracted 105{,}163 fixed 10\,s windows to form the ``WhaleSounds'' dataset; Dempster et al.\ \cite{dempster2024highly} used it as a large-scale TSC benchmark. However, 10 s fixed windows truncate 20--40\,s Whale calls, random CV discards hierarchical site-year structure, isolated-call extraction removes negative-class soundscape diversity, and accuracy alone is reported—limiting relevance to real UPAM detection difficulty.

\paragraph{Evaluation Protocols in UPAM}
UPAM evaluation is complicated by spatial and temporal dependence: standard CV often ignores these structures \cite{Valavi2019, Roberts2017, Racine2000}, inflating performance when train/test sets share site- or year-specific traits and causing failures in new environments \cite{Hobday2019,Young2022}. Annotation quality also varies: Dubus et al.\ \cite{dubus2024citizen} reannotated Elephant Island 2013 and found substantial expert–novice disagreement, consistent with known inconsistencies at this site \cite{Miller2021}, motivating its exclusion here. Schall et al.\ \cite{Schall2024} proposed a site-year blocked benchmark reporting TCR, NMR, CMR, and Overall Fitness, but macro-averaged metrics can mask poor performance on rare calls, omit AP/Precision/Recall/F1, and provide no fold-level stability. Although nested CV is used elsewhere for ensembling \cite{Roberts2017} or hyperparameter tuning \cite{Schratz2019}, these uses do not expose within-block variance—critical in UPAM, where heterogeneity and class imbalance make stability essential for deployment readiness and conservation estimates.

\paragraph{Our Benchmark GetNetUPAM} 
We extend blocked CV with a nested layer \emph{within} each site-year block, using the nested stage to \textbf{quantify stability} rather than inflate hold-out scores. Multiple tests on individual hold-outs yield a stability profile complementing mean metrics. By combining macro/micro scores with AP, Recall, Precision, and F1, GetNetUPAM provides a reproducible, ecologically grounded evaluation reflecting deployment-like conditions. Given dataset scale where a 1\% false-positive rate yields thousands of spurious detections, rigorous stability assessment is essential for operational reliability for clear estimates \cite{Saito2015}.

\paragraph{Our Neural Architecture ARPA-N} 
CBAM’s \cite{woo2018cbam} spatial mapping suppresses non-informative spectrogram regions and improves robustness under heterogeneous noise, as shown in multi-resolution attention for environmental sound classification \cite{Zhang2022} and CBAM-augmented CNNs show stability down to $-$20\,dB SNR by avoiding global spectral shortcuts \cite{wang2024spectrum, geirhos2020shortcut}. These behaviors parallel long-window UPAM failure modes, where non-attention CNNs exploit global coloration rather than call structure. Placement studies show early/mid-layer attention generalizes best, while full-stack attention can overfit \cite{huai2025heart}. Motivated by these findings and the strong spatial-dropout effects in ARP-N, ARPA-N integrates adaptive resolution pooling with strategically placed CBAM spatial attention to suppress localized noise while preserving diagnostically relevant structure and efficiency, directly addressing shortcut learning and underpinning the cross-site stability and performance gains observed in our experiments.

\section{Method}
\label{Deep}

\begin{figure}[t!]
    \centering
    \includegraphics[width=0.8\linewidth]{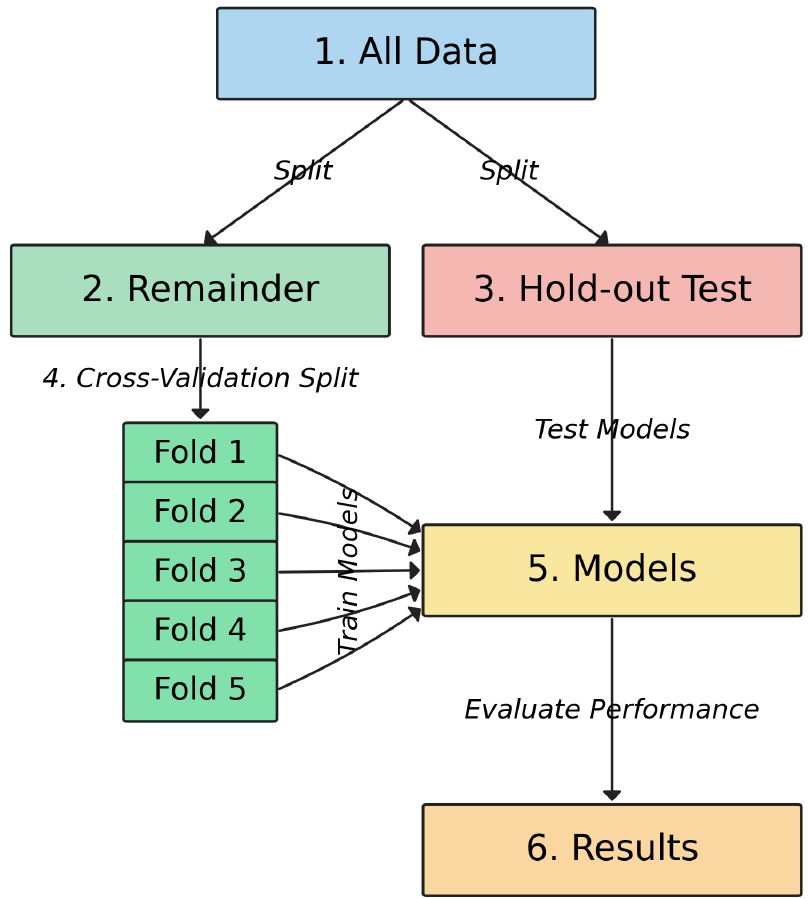}
    \caption{Overview of the hierarchical nested cross-validation architecture, from data partitioning and model training to evaluation. 
    The nested structure enforces strict separation between development and evaluation, yielding fold-level stability profiles that reflect deployment-like environmental variability.}
    \label{fig:testDiagram}
\end{figure}

\begin{figure*}[t!]
\centering
\includegraphics[width=\linewidth]{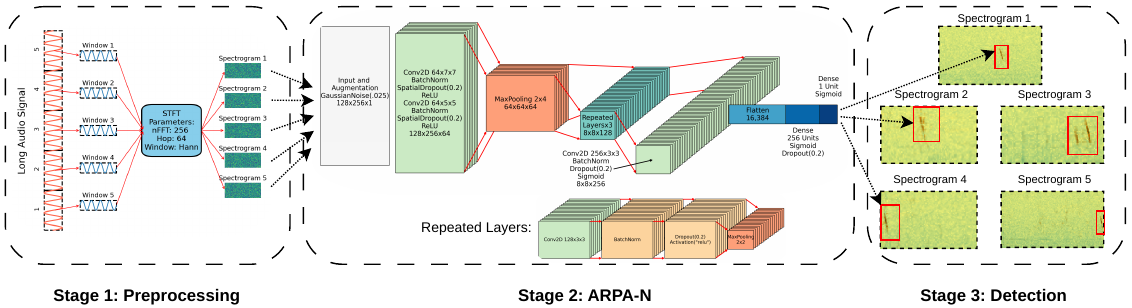} 
\caption{Integrated preprocessing and ARPA‑N detection pipeline. Stage 1: raw audio is windowed and converted to spectrograms via STFT. Stage 2: ARPA‑N processes them through initial convolutions, adaptive pooling, and spatial‑attention blocks. Stage 3: final detection outputs are produced.}
\label{fig:model}
\end{figure*}

\textbf{GetNetUPAM} is a benchmarking framework for evaluating the generalization and computational efficiency of neural networks in Underwater Passive Acoustic Monitoring (UPAM). It integrates five components: hierarchical nested cross-validation, windowing, time–frequency transformation, efficient model design, and detection; thus, providing a stability‑aware assessment of \textbf{ARPA‑N}, our lightweight attention‑based CNN for irregular spectrograms. The nested cross‑validation procedure is formalized in Algorithm~\ref{Alg:GNU} and illustrated in Figure~\ref{fig:testDiagram}, with the full ARPA‑N pipeline shown in Figure~\ref{fig:model}, and Figure~\ref{fig:spAttn} visualizing CBAM's attention modules.

\begin{algorithm}[t!]
\caption{GetNetUPAM: Nested Cross-Validation }
\label{Alg:GNU}
\begin{algorithmic}
\small
\REQUIRE Dataset $D$, \# outer folds $K$, \# inner folds $k$
\ENSURE Mean and standard deviation of metrics
\STATE Split $D$ into $K$ site-years $\{D_1, D_2, \ldots, D_K\}$
\FOR{$i \gets 1$ to $K$}
    \STATE Assign $D_i$ as the outer test set
    \STATE Combine remaining data into training set $T_i$
    \STATE Split $T_i$ into $k$ inner folds $\{T_{i1}, \ldots, T_{ik}\}$
    \FOR{$j \gets 1$ to $k$}
        \STATE Assign $T_{ij}^{\text{val}}$ as the inner validation set
        \STATE Combine remaining folds into $T_{ij}^{\text{train}}$
        \STATE Train model $M_{ij}$ on $T_{ij}^{\text{train}}$
        \STATE Validate $M_{ij}$ on $T_{ij}^{\text{val}}$
        \STATE Test $M_{ij}$ on $D_i$
    \ENDFOR
    \STATE Compute mean and std.\ of test metrics for $D_i$
\ENDFOR
\STATE Compute micro- and macro-averaged metrics across all $D_i$
\end{algorithmic}
\end{algorithm}

\subsection{GetNetUPAM: Hierarchical Nested Cross‑Validation}
\label{GetNet}

UPAM datasets exhibit strong spatial and temporal dependencies, making random cross‑validation unreliable for estimating real‑world performance \cite{Roberts2017}. Such splits risk training–test leakage when site‑ or year‑specific patterns overlap, inflating accuracy and undermining deployment reliability. GetNetUPAM addresses this with a two‑level \emph{hierarchical nested cross‑validation} scheme.

In the \textbf{outer loop}, each site‑year is held out as the test set, ensuring evaluation under unseen spatiotemporal conditions. The remaining site‑years form the training pool. Within this pool, the \textbf{inner loop} performs five‑fold stratified cross‑validation, preserving class balance. For each outer test set, five models are trained on the inner folds and evaluated on the same held‑out site‑year, producing a distribution of scores. Mean and standard deviation from this distribution directly quantify model stability under deployment‑like variability.

To complement predictive performance and stability we also evaluate efficiency with benchmark inference speed using the full Balleny Island 2015 site‑year \cite{Miller2021} and model parameter counts. Thus, we measure total and per‑sample inference time and total trainable parameters, ensuring models are both precise and feasible for large‑scale, long‑term monitoring. The full experimental setup is detailed in Section~\ref{Res}.

\subsection{Windowing}
\label{win}

Stage one segments continuous audio into fixed-length, overlapping windows, a standard approach for sequential data in deep neural networks \cite{McCallum2018}. The raw waveform, sampled at \(250\)~Hz, is divided into 16,384-sample segments (65.536~seconds), balancing temporal resolution with the likelihood of capturing multiple complete vocalizations \cite{Rasmussen2024}.

We apply a sliding window with 50\% overlap, defined as \(w = 2 \times h\), where \(w\) is the window size and \(h\) is the hop size. This overlap mitigates boundary effects by ensuring calls spanning segment edges appear fully in at least one window, while improving the balance between positive and negative samples. The total number of windows generated from an audio file of length \(s\) samples is: \(
N = \left\lfloor \frac{s - h}{h} \right\rfloor,
\)where the floor operation discards any truncated final segment, and \(s-h\) accounts for the 50\% overlap.

\subsection{2D Time–Frequency Data Representation}
\label{stft}

Building on the segmentation in Section~\ref{win}, each 65.536\,s audio window (16{,}384 samples at 250\,Hz) is transformed into a 2D time–frequency representation via the Short-Time Fourier Transform (STFT). Parameters are matched to the windowed segments: window length $L$ and hop size $b$ preserve temporal granularity while enabling fine spectral resolution. This alignment ensures the temporal context from windowing is retained in the spectral domain, allowing the network to learn jointly from time structure and frequency content \cite{Cordero2025}. This projection reveals biologically informative patterns, e.g., the 20–100\,Hz band of blue whale anthems or the 40–120\,kHz range of dolphin clicks, in a form exploitable by convolutional architectures. Such hierarchical encoding over frequency and time helps model multi‑species vocalizations.

In a discrete-time signal $\mathbf{w}[n] \in \mathbb{R}^T$, the STFT is:
\begin{equation}
\mathbf{STFT}(i,f) = \sum_{k=0}^{L-1} \mathbf{w}[bi + k] \cdot \mathbf{x}[k] \cdot e^{-j2\pi kf/L},
\label{eq:stft}
\end{equation}
where $L$ is the window length, $b$ the hop size, $\mathbf{x}[k]$ the window function (e.g., Hann), $f$ the frequency bin index, and $i$ the time frame index. This yields a complex-valued $N \times L$ matrix, with $N = \lfloor T / b \rfloor$ and frequency resolution $\Delta f = f_s / L$ for sampling rate $f_s$. Discarding the redundant half of the spectrum (Hermitian symmetry) yields $M \times N$ with $M = L/2$. For $L=256$ and $b=64$, we obtain $128 \times 256$ spectrograms, each column representing a 256-sample segment spaced 64 samples apart (75\% overlap). After we convert to a log-power spectrogram: \(
\mathbf{P}_{\text{dB}}(i,f) = 10 \cdot \log_{10} \left( \left| \mathbf{STFT}(i,f) \right|^2 + \epsilon \right),
\) as $\epsilon$ ensures numerical stability, then apply min–max scaling: \(
\mathbf{P}_{\text{norm}}(i,f) = \frac{\mathbf{P}_{\text{dB}}(i,f) - \min_{i,f} \mathbf{P}_{\text{dB}}}{\max_{i,f} \mathbf{P}_{\text{dB}} - \min_{i,f} \mathbf{P}_{\text{dB}}}.
\)

Each column corresponds to $[bi, bi + L)$ in the waveform. A convolutional kernel spanning $k_t$ frames has receptive field:
\begin{equation}
R_{\text{time}} = L + (k_t - 1) \cdot b,
\end{equation}
e.g., $k_t = 5$ gives $R_{\text{time}} = 512$ samples, extending temporal context without explicit long‑range attention.

Convolutional filters $W \in \mathbb{R}^{k_t \times k_f}$ operate locally:
\begin{equation}
G(i,f) = \sum_{p=0}^{k_t - 1} \sum_{q=0}^{k_f - 1} W(p,q) \cdot \mathbf{P}_{\text{norm}}(i+p, f+q),
\label{EQ:convR}
\end{equation}
aggregating localized patterns such as harmonics, shifts, and transients. Overlapping STFT windows give multiple, slightly shifted views of each region, capturing transitions that non‑overlapping transformer tokenization may miss. While transformers offer global context, redundancy enriches local representations, providing robust complex underwater signals.

\subsection{Adaptive Resolution Pooling and Attention Network}
\label{ARPA-N}

Stage two takes the normalized log-power spectrograms from Sections~\ref{win}–\ref{stft} and feeds them into \textbf{ARPA-N}, a lightweight attention-based CNN built to handle the odd spatial dimensions resulting from whale call spectrograms. These arise directly from earlier segmentation and STFT choices; ARPA-N’s first role is to reconcile them with the backbone.

Architecturally, ARPA-N follows VGG16’s use of small $3\times3$ convolutions and max pooling, adapted for spectral data and our input geometry. Key refinements include:  
\textbf{Reduced depth between pooling} — one convolution before each pooling step, controlling complexity, \textbf{Adaptive pooling for odd dimensions} — early layers reshape spectrograms to match the backbone to standardize feature maps for stability and scalability, and \textbf{Spatial attention integration} — CBAM’s spatial attention \cite{woo2018cbam} expands the receptive field without larger kernels, focusing on salient spectro‑temporal regions while reducing noise. This combination yields a model robust to large, heterogeneous datasets yet sensitive to spectral cues.

\paragraph{Initial processing and attention.}
The first layer applies Additive Gaussian Noise:
\(
\mathbf{O}^{(0)} = \mathbf{P_{\text{norm}}} + |\mathcal{N}(0, \sigma^2)|,
\)
followed by a $7\times7$ convolution with $64$ filters and 'same' padding: \(
\mathbf{O}^{(l)} = \mathbf{O}^{(l-1)} \circledast \mathbf{W}^{(l)} + \mathbf{b}^{(l)},
\) batch normalization  
\( \mathbf{\hat{O}^{(l)}_i} = \frac{\mathbf{O^{(l)}_i} - \mu_{\text{batch}}}{\sqrt{\sigma_{\text{batch}}^2 + \epsilon}} \). With spatial dropout (probability $p = .2$) \cite{Rasmussen2023}, and ReLU \( \tilde{\mathbf{O}}^{(l)} = \phi\!\left( \mathbf{\hat{O}}^{(l)} \odot \mathbf{M}^{(l)} \right) \) afterward.

\begin{figure}[t!] \centering \includegraphics[width=1.0\linewidth]{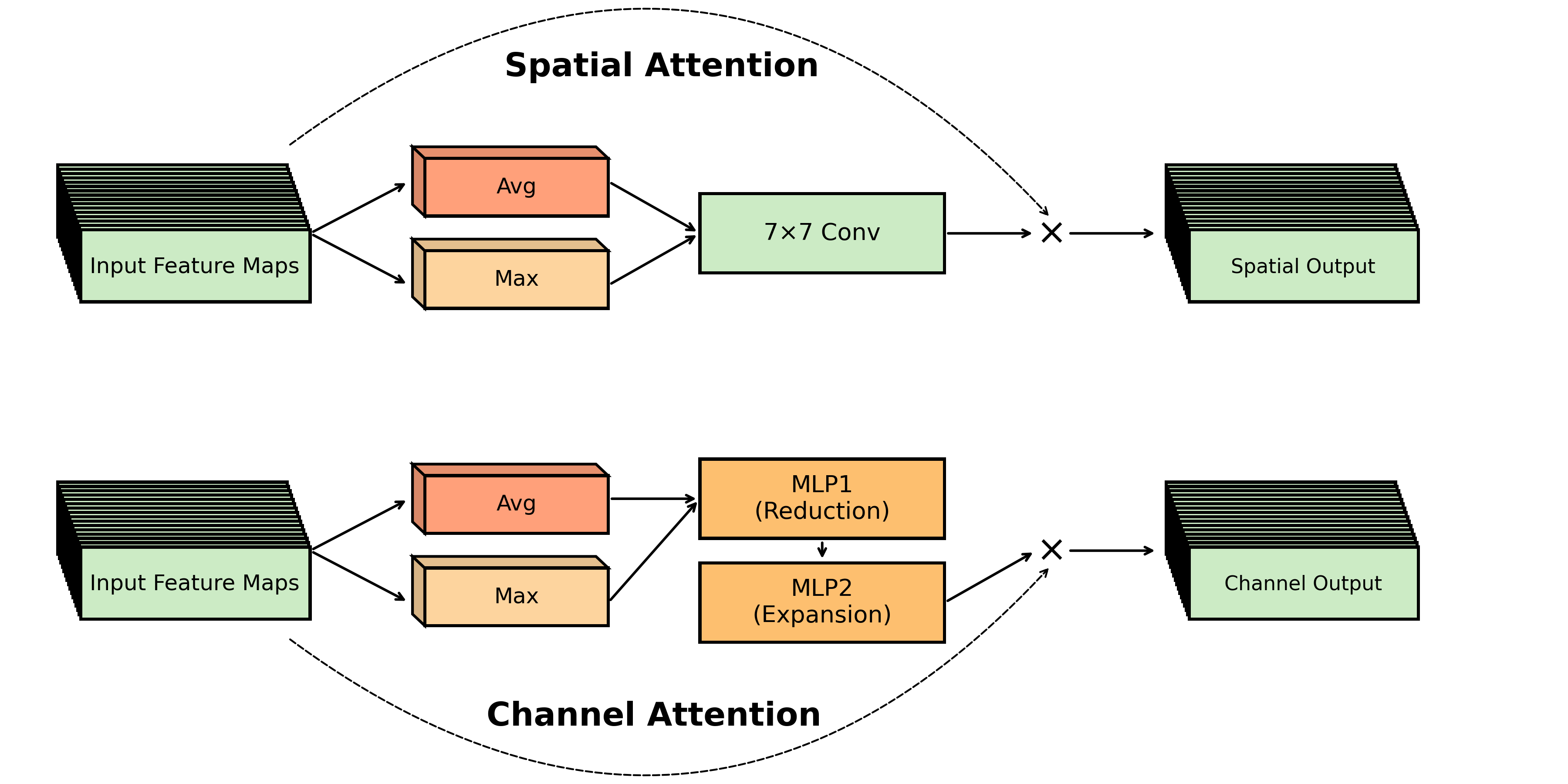} \caption{\textbf{Spatial Attention} pools the input along the channel dimension (average and max), concatenates the resulting 2‑D maps, and applies a 7×7 convolution to generate a spatial‑attention map that is broadcast and multiplied with the input. \textbf{Channel Attention} applies average‑ and max‑pooling in parallel for each feature map, feeds the pooled descriptors through a shared MLP, and produces a channel‑attention vector that reweights the feature maps.} \label{fig:spAttn} \end{figure}

We then place these outputs into CBAM's spatial attention \cite{woo2018cbam}, as in Figure \ref{fig:spAttn}, which refines features. Given \(\tilde{\mathbf{O}}^{(l)} \in \mathbb{R}^{H \times W \times C}\), we compute average and max descriptors of inputs:
\begin{equation}
\mathbf{O}_{\text{avg}}(i,j) = \frac{1}{C} \sum_{k=1}^{C} \tilde{\mathbf{O}}^{(l)}(i,j,k),
\end{equation}
\begin{equation}
\mathbf{O}_{\text{max}}(i,j) = \max_{1 \leq k \leq C} \tilde{\mathbf{O}}^{(l)}(i,j,k),
\end{equation}
concatenate them into \(\mathbf{O}_{\text{cat}}\), and convolve a $7\times7$ kernel \(\mathbf{K}\):
\(
\dot{\mathbf{O}}(i,j) = \sum_{m=-3}^{3} \sum_{n=-3}^{3} \mathbf{K}(m,n) \cdot \mathbf{O}_{\text{cat}}(i+m, j+n).
\) Batch normalization and a sigmoid produce the attention map  
\(\mathbf{M}_{\text{spatial}}(i,j) = \sigma(\dot{\mathbf{O}}_{\text{BN}}(i,j))\),  
broadcast across channels and applied element‑wise:  
\(\ddot{\mathbf{O}}^{(l)}(i,j,k) = \tilde{\mathbf{O}}^{(l)}(i,j,k) \cdot \mathbf{M}_{\text{spatial}}(i,j)\).  
We repeat the procedure with a $5\times5$ initial convolution. As shown in ablation~\ref{tab:ab}, CBAM channel attention consistently reduced cross‑site stability and lowered precision, reinforcing global spectral shortcuts rather than suppressing them. Because it also increases inference cost in multi‑path variants without improving FP/hr$_{90}$, ARPA‑N omits channel attention and retains only the spatial module, learning to suppress noise.

\paragraph{Adaptive resolution pooling.}
After the convolution–attention stages, a $2\times4$ max pooling reduces height and width by 2 and 4:
\(
\text{Height}_{\text{out}} = \frac{\text{Height}_{\text{in}}}{2}, \quad
\text{Width}_{\text{out}} = \frac{\text{Width}_{\text{in}}}{4},
\)
an $8\times$ area reduction. Outputs are standardized to:
\begin{equation}
\text{Height}_{\text{out}} = \text{Width}_{\text{out}} = \text{Channels}_{\text{out}} = 64,
\end{equation}
cutting computation and enabling transfer learning across varying spectrogram resolutions. By linking adaptive pooling to the STFT‑derived odd‑dimension spectrograms, ARPA‑N bridges raw acoustic structure and standardized neural classification architecture, retaining efficiency without sacrificing sensitivity to biologically relevant features.

\paragraph{Deeper feature extraction.}
Three $3\times3$ convolution layers (128 filters) with Batch Norm, ReLU, and $2\times2$ max pooling further condense features. A final $3\times3$ convolution (256 filters) precedes sigmoid activation
\(
\tilde{\mathbf{O}}^{(-1)} = \sigma \left( \mathbf{\hat{O}}^{(-1)} \odot \mathbf{M}^{(-1)} \right),
\)
limiting dynamic range \cite{Dubey2022} yet keeping critical information \cite{Braun2010}. Then the output is flattened: \( w = |flatten(\ddot{\mathbf{O}}^{(-1)}(i,j,k))|,
\) preserving spatial relationships.

\begin{table*}[t]
\scriptsize
\centering
\renewcommand{\arraystretch}{1.3}
\setlength{\tabcolsep}{4pt}

\begin{tabular}{|l|ccccccccccc|}
\toprule \rowcolor{gray!35}
\textbf{Metric} &
\textbf{1(T)} & \textbf{2(X)} & \textbf{3} & \textbf{4} & \textbf{5} &
\textbf{6} & \textbf{7} & \textbf{8(T)} & \textbf{9} & \textbf{10} & \textbf{11(T)} \\
\midrule \rowcolor{gray!15}
\multicolumn{12}{c}{\textbf{60-second Windows}} \\
\midrule
Windows &
22040 & 67864 & 20736 & 3230 & 8360 &
19008 & 20940 & 20187 & 21600 & 21578 & 21591 \\

Call Windows &
67 & 5743 & 1091 & 90 & 87 &
0 & 967 & 884 & 458 & 697 & 1306 \\

Ratio &
.0030 & .0846 & .0526 & .0279 & .0104 &
.0000 & .0462 & .0438 & .0212 & .0323 & .0605 \\
\midrule \rowcolor{gray!15}
\multicolumn{12}{c}{\textbf{4.5-second Windows}} \\
\midrule
Windows &
293511 & 1009432 & 308427 & 45600 & 113810 &
253215 & 279002 & 268956 & 287800 & 287507 & 287675 \\

Call Windows &
155 & 22679 & 3249 & 190 & 231 &
0 & 1716 & 1572 & 1088 & 1205 & 3095 \\

Ratio &
.0005 & .0225 & .0105 & .0042 & .0020 &
.0000 & .0062 & .0058 & .0038 & .0042 & .0108 \\
\bottomrule
\end{tabular}
\caption{Dataset availability across the eleven Antarctic site–year deployments (Datasets~1–11): (1)~BallenyIslands2015\,(T), (2)~ElephantIsland2013Aural\,(X), (3)~ElephantIsland2014, (4)~Greenwich64S2015, (5)~MaudRise2014, (6)~RossSea2014, (7)~Casey2014, (8)~Casey2017\,(T), (9)~Kerguelen2005, (10)~Kerguelen2014, and (11)~Kerguelen2015\,(T). BallenyIslands2015, Casey2017, and Kerguelen2015 are held‑out test sets; ElephantIsland2013Aural is excluded (X) due to annotation issues. For each dataset, we report total windows, call‑positive windows, and call–window ratios under 60‑s and 4.5‑s windowing.}
\label{tab:dataset_columns_windowing}
\end{table*}

\subsection{Detection}

Stage three takes the flattened feature vector $w$ output by ARPA-N (Section~\ref{ARPA-N}) and passes it through a lightweight multi-layer perceptron to produce class likelihoods. A 256-unit dense layer first transforms the features:
\(
\mathbf{D} = \sigma(\mathbf{W}_d\,w + \mathbf{b}_d),
\)
followed by an output layer yielding per-class probabilities
\(
\hat{y}_i = \sigma(\mathbf{W}_o\,\mathbf{D} + \mathbf{b}_o),
\)
where $\hat{y}_i$ is the probability of class $i$.

\paragraph{Saliency-guided time--frequency highlighting.}
To visualize which regions of the input spectrogram most influenced the model's prediction, we generate class-specific saliency maps~\cite{Zhou2016} by computing the gradient of the predicted class score with respect to the normalized STFT input.
For an input tensor $\mathbf{P}_{\text{norm}}(i,f)$ and target class $k$, the saliency value at time--frequency bin $(i,f)$ is defined as
\begin{equation}
S(i,f) = \left| \frac{\partial \hat{y}_k}{\partial \mathbf{P}_{\text{norm}}(i,f)} \right|,
\end{equation}
yielding interpretable time--frequency regions that most influence the decision.

\paragraph{Embedding analysis.}
To characterize the model’s internal representation, we extract the 256-dimensional activation \( \mathbf{D} \) from the penultimate dense layer. This vector provides a compact embedding of each input window, capturing the discriminative time--frequency patterns emphasized by ARPA-N. We aggregate embeddings across the dataset and examine their structure using PCA and t\textendash SNE/UMAP to assess class separability and to identify clusters reflecting recurrent acoustic motifs. Because \( \mathbf{D} \) precedes the final decision layer, it preserves the geometry of the learned feature space and supports stable visualization and error analysis.

\section{Experimental Setup and Results}
\label{Res}

This section details the experimental setup and evaluation framework for GetNetUPAM, including datasets, metrics, computational resources, and baseline architectures. We then present core results, linking outcomes to the methodological components in Section~\ref{Deep} to show how hierarchical nested cross-validation, ARPA-N, and the detection pipeline contribute to performance. Quantitative analysis is complemented by qualitative inspection of samples from our detection algorithm, followed by an ablation study isolating network components, reflecting the modular design in the Methods section \ref{Deep}.

\begin{figure}[t!]
\centering
\includegraphics[width=1.00\linewidth]{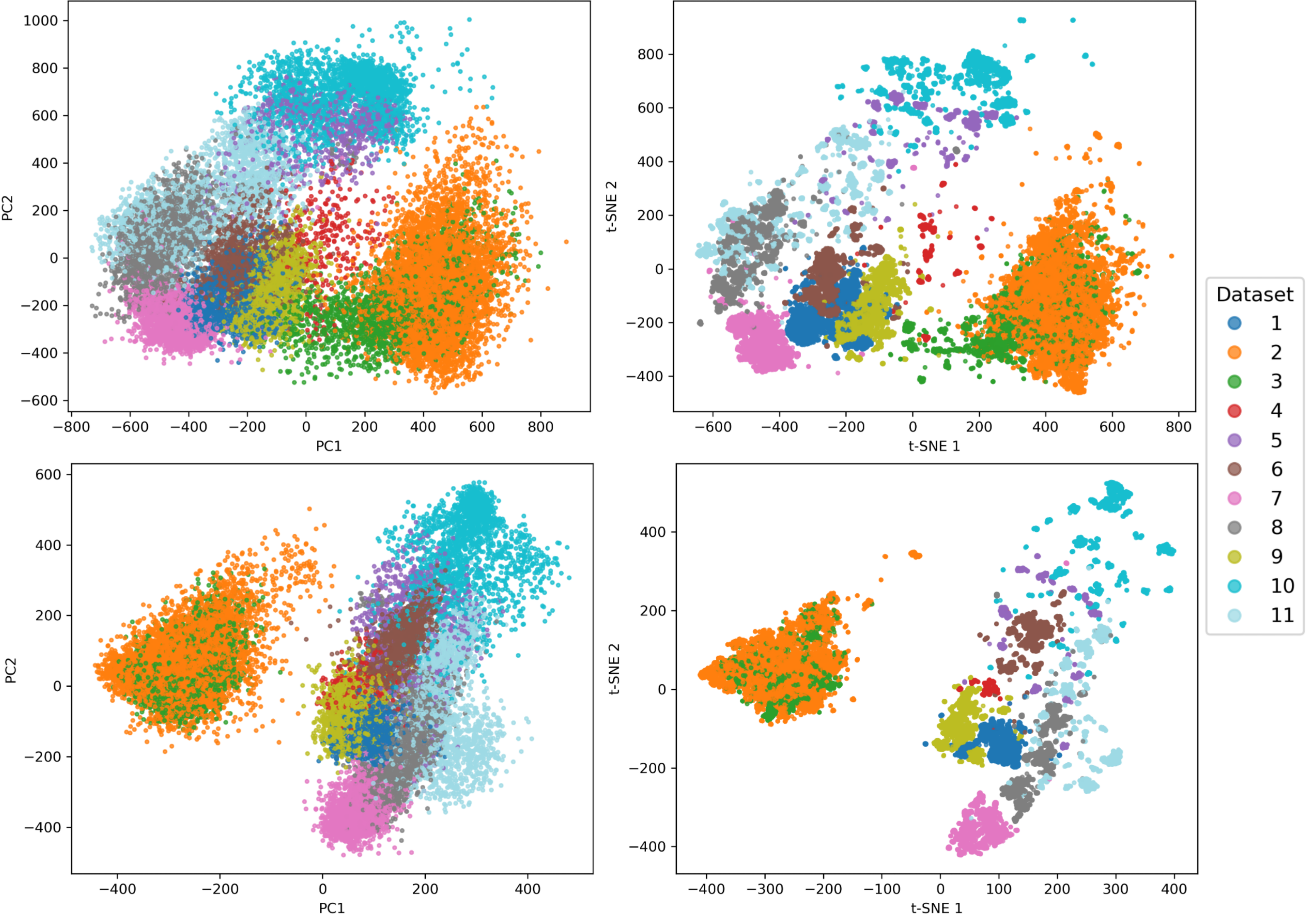} 
\caption{Low‑dimensional projections of the learned embeddings for both model variants. Left: PCA; right: t‑SNE. Top row: APR‑N; bottom row: ARPA‑N. Each point is a 16{,}384‑sample window colored by site–year, with both projections computed from the same embeddings. PCA shows dominant linear variance, while t‑SNE highlights local nonlinear structure, revealing how APR‑N and ARPA‑N organize the acoustic feature space across deployments.}
\label{fig:PRC}
\end{figure}

\subsection{Dataset Quantification and Separability Analysis}
\label{dataset}

Blue Whale D‑calls are sparse and inconsistently annotated, yet remain demographically important for estimating female abundance \cite{Miller2022}. We use Kerguelen~2015, Casey~2017, and Balleny Islands~2015 from the ATBFL \cite{Miller2021} as held‑out test sets, spanning a wide range of positive samples (1180, 553, and 47) and differing levels of supporting training data (two, one, and zero years). This variation provides a natural gradient for evaluating GetNetUPAM’s blocked site‑year generalization (Section~\ref{GetNet}) WRT positive class availability.

Elephant Islands~2013 \cite{Miller2021} was removed due to inconsistencies between its documentation and annotator notes, especially in FM‑call labeling. Elephant Islands~2014 was included for training but excluded from testing to avoid bias, mirroring our control of training/test leakage in the hierarchical nested CV.

Table~\ref{tab:dataset_columns_windowing} summarizes availability across the eleven Antarctic site–years (Datasets~1–11), reporting total windows, call‑positive windows, and call–window ratios under 60‑s and 4.5‑s windowing. Long 60‑s windows yield 3{,}230–67{,}864 samples with call ratios up to 8.46\%, as extended context increases the likelihood of capturing a D‑call and smooths local variability, inflating baseline separability. Short 4.5‑s windows increase sample counts by an order of magnitude (45{,}600–1{,}009{,}432) but reduce call ratios to 0.05\%–2.25\%, isolating calls while greatly expanding the non‑call class. Thus, long windows emphasize global acoustic context—including site‑level coloration and annotator biases—whereas short windows suppress these cues and force reliance on localized call structure, increasing class imbalance and task difficulty.

Figure~\ref{fig:PRC} provides embedding visualizations for how this window‑driven separability interacts with architectural bias in a multi-class dataset classification experiment. ARP‑N converges rapidly on 60‑s data (loss $<0.07$, accuracy $\sim97.6\%$, AUC $\sim0.999$), indicating reliance on global, non‑biological cues such as noise floors and hydrophone characteristics. ARPA‑N shows lower early accuracy ($\sim70\%$), fluctuating AUC, slower convergence, and higher final loss ($\sim0.185$), with best‑epoch accuracy of 93.8\% and AUC 0.995. These dynamics reflect spatial attention suppressing global shortcuts and enforcing localized feature use, leading to repeated restructuring of the embedding space before stabilizing. Although ARPA‑N ultimately learns a strong ranking function, its decision boundary differs fundamentally from ARP‑N’s shortcut‑driven one. Overall, long windows favor models exploiting global structure, while spatial‑attention models perform worse on dataset separability precisely because they suppress these cues showing that global context can be unreliable and the true signal is inherently local, further explaining baseline ViT results.

\subsection{Evaluation}

\paragraph{Evaluation Metrics} We employ AP, recall, precision, and F1-score. Additionally, we compute FP/hr by computing different levels of recall on frames-level predictions. Most critically for deployment readiness, we measure model stability via standard deviation:
\(
\sigma = \sqrt{\frac{1}{n} \sum_{i=1}^{n} (x_i - \mu)^2},
\)
where $x_i$ are individual metrics, $\mu$ is their mean, and $n$ the number of folds. This stability metric directly reflects the variability of training in the nested inner loop in Section~\ref{GetNet}.

We also compute Micro and Macro averages:
\begin{equation}
\label{eq:Mic}
\text{Micro} = \frac{\sum_{i=0}^{n} M_i \times N_i}{\sum_{i=0}^{n} N_i}, \quad
\text{Macro} = \frac{\sum_{i=0}^{n} M_i}{n},
\end{equation}
where $M_i$ is the metric for dataset $i$, $N_i$ its annotation count, and $n$ the number of datasets. With class imbalance up to 170:1, negative-class metrics (accuracy, specificity, ROC-AUC) are omitted to avoid bias. For consistency with Section~\ref{win}, we concatenate up to three adjacent positives for the 60-second variant and unlimited positives for the 4-second variant.

\paragraph{Experimental Parameters} At each GetNetUPAM training iteration, negative windows are downsampled by half, using the 50\% temporal overlap in Section~\ref{win} to preserve acoustic diversity while controlling class imbalance. Except for pre‑trained baselines, all models are trained from scratch to capture whale‑specific spectral–temporal structure \cite{alzubaidi2021deepening,Guzhov2021}. Training uses binary cross‑entropy and the Adam optimizer (initial LR 0.01, halved every five epochs) \cite{Kingma2015}, balancing convergence speed with the stability demands of nested CV. Model selection is based on highest validation accuracy, after which the five chosen weights are evaluated on the held‑out site‑year in the outer loop, producing stability estimates.

\paragraph{Computational Resources} Experiments ran on the Anonymous Anonymous Anonymous Computing Center (AAAC) (NSF Grant Anonymous) within a Slurm environment. We used 24 CPUs and 160\,GB RAM; most training ran on NVIDIA V100 32\,GB GPUs, with P100 16\,GB GPUs used for efficiency measurements in Tables~\ref{tab:res} and~\ref{tab:ab}. This dual-GPU setup separated inference efficiency from training throughput, ensuring metrics match deployment conditions. Code will be made available upon acceptance; reviewers may request access during the review process.

\begin{table*}[t!]
\scriptsize
\centering
\begin{tabularx}{\textwidth}{|>{\columncolor{gray!15}}l|XXXXXXXl|XXXXXXXl|XXXXXXXl|}
\toprule
\rowcolor{gray!35}
\multicolumn{1}{|c}{} & \multicolumn{4}{|c}{\textbf{Kerguelen 2015}} & \multicolumn{4}{c|}{\textbf{2 Years Support}} &  \multicolumn{4}{c}{\textbf{Casey 2017}} & \multicolumn{4}{c|}{\textbf{1 Year Support}} & \multicolumn{4}{c}{\textbf{BallenyIslands 2015}} & \multicolumn{4}{c|}{\textbf{No Support}} \\ \midrule
\rowcolor{gray!15}
\textbf{Model} & \textbf{AP} & $\sigma$ & \textbf{Rec} & $\sigma$ & \textbf{Pre} & $\sigma$ & \textbf{F1} & $\sigma$ & \textbf{AP} & $\sigma$ & \textbf{Rec} & $\sigma$ & \textbf{Pre} & $\sigma$ & \textbf{F1} & $\sigma$ & \textbf{AP} & $\sigma$ & \textbf{Rec} & $\sigma$ & \textbf{Pre} & $\sigma$ & \textbf{F1} & $\sigma$ \\ 
\midrule

Base\cite{Miller2022} & .361 & .025 & .460 & .084 & .795 & .069 & .575 & .060 & .200 & .035 & .450 & .149 & .419 & .168 & .391 & .054 & .094 & .008 & .430 & .175 & \textbf{.256} & .052 & .288 & .053 \\
4sRes & .370 & .033 & .141 & .087 & \textbf{.908} & .054 & .232 & .114 & .196 & .037 & .284 & .109 & .563 & .203 & .332 & .087 & .089 & .010 & .106 & .082 & \textbf{.468} & .267 & .143 & .104 \\
4sSCNN & .382 & .015 & .241 & .047 & \textbf{.877} & \textbf{.019} & .375 & .057 & .223 & .024 & .365 & .034 & .503 & .075 & .418 & .029 & .086 & .007 & .179 & .079 & \textbf{.268} & .029 & .204 & .061 \\
4sARP-N & .413 & .012 & .615 & .079 & .741 & .048 & .666 & .036 & .247 & .030 & .684 & .028 & .283 & .080 & .392 & .069 & .107 & .009 & .528 & .064 & \textbf{.259} & .034 & .346 & .040 \\
4sResA & .465 & .082 & .564 & .093 & .756 & .046 & .638 & .057 & .399 & .022 & .579 & .044 & .537 & .055 & .553 & .023 & .142 & .019 & .549 & .124 & .197 & .020 & .285 & .028 \\
4sARP-NA & .564 & .013 & .652 & .094 & .707 & .070 & .669 & .024 & .414 & .026 & .718 & .037 & .264 & .032 & .385 & .035 & .158 & .008 & .562 & .113 & .189 & .029 & .275 & .016 \\
60sPre & .677 & .032 & .789 & .035 & .734 & .052 & .759 & .022 & .337 & .167 & .462 & .173 & .414 & .157 & .435 & .160 & .252 & .038 & \textbf{.877} & .053 & .165 & .047 & .273 & .065 \\
60sBase\cite{Rasmussen2024} & .720 & .017 & .794 & .026 & .807 & .032 & .799 & .006 & .580 & .016 & .613 & .013 & .693 & .036 & .650 & .020 & .282 & .023 & .766 & .038 & \textbf{.248} & .020 & .374 & .020 \\
60sDense & .791 & .024 & .857 & .020 & .819 & .023 & .837 & .011 & .627 & .024 & .698 & .036 & .678 & .036 & .687 & .019 & .298 & .007 & .702 & .043 & \textbf{.389} & .069 & \textbf{.495} & .052 \\
ARP-N & \textbf{.850} & \textbf{.005} & \textbf{.903} & \textbf{.017} & .794 & .019 & .845 & .004 & .710 & .005 & \textbf{.790} & .039 & .648 & .044 & .709 & .011 & .330 & .030 & .821 & .022 & \textbf{.235} & .028 & .364 & .033 \\
ARPA-FN & \textbf{.852} & .006 & .883 & .023 & .812 & .021 & .845 & .007 & .722 & .009 & \textbf{.756} & \textbf{.018} & .697 & .019 & .725 & .007 & .309 & .011 & \textbf{.838} & \textbf{.017} & \textbf{.267} & \textbf{.014} & .404 & .014 \\
ARPA-N & \textbf{.857} & .008 & .823 & .023 & \textbf{.888} & .023 & \textbf{.854} & \textbf{.003} & \textbf{.744} & \textbf{.008} & .713 & .013 & \textbf{.756} & \textbf{.016} & \textbf{.733} & \textbf{.004} & \textbf{.367} & \textbf{.017} & .706 & .045 & \textbf{.357} & .019 & \textbf{.474} & \textbf{.023} \\

\bottomrule
\end{tabularx}

\begin{tabularx}{\textwidth}{|>{\columncolor{gray!15}}l|XXXXXXXl|XXXXXXXl|XXXXXXXl|}
\toprule
\rowcolor{gray!15}
\textbf{Model} & \textbf{R75} & $\sigma$ & \textbf{R80} & $\sigma$ & \textbf{R85} & $\sigma$ & \textbf{R90} & $\sigma$ & \textbf{R75} & $\sigma$ & \textbf{R80} & $\sigma$ & \textbf{R85} & $\sigma$ & \textbf{R90} & $\sigma$ & \textbf{R75} & $\sigma$ & \textbf{R80} & $\sigma$ & \textbf{R85} & $\sigma$ & \textbf{R90} & $\sigma$ \\ 
\midrule

4sBase\cite{Miller2022} & 511 & 77.7 & 681 & 75.4 & 859 & 82.8 & 1040 & 68.5 & 385 & 39.1 & 563 & 53 & 756 & 51.9 & 972 & 45.7 & 623 & 181 & 725 & 184 & 926 & 175 & 1090 & 154 \\
4sRes & 381 & 111 & 533 & 124 & 698 & 136 & 898 & 118 & 338 & 36.6 & 503 & 64.2 & 702 & 55.7 & 933 & 53.3 & 540 & 235 & 659 & 232 & 835 & 236 & 1010 & 217 \\
4sSCNN & 229 & 28.9 & 362 & 37.2 & 553 & 34.5 & 779 & 38.2 & 339 & 71.8 & 520 & 98.1 & 708 & 102 & 945 & 105 & 408 & 62.7 & 542 & 114 & 745 & 106 & 942 & 80.8 \\
4sARP-N & 285 & 39.6 & 456 & 66.1 & 656 & 84.7 & 888 & 99.1 & 302 & 24.3 & 479 & 38.7 & 677 & 49.5 & 889 & 23.6 & 479 & 76.7 & 624 & 102 & 770 & 104 & 997 & 65.8 \\
4sResA & 292 & 191 & 422 & 184 & 594 & 162 & 780 & 137 & 247 & 59.2 & 407 & 76.2 & 614 & 83.2 & 821 & 76.6 & 208 & 51.5 & 272 & 54.8 & 380 & 69.6 & 486 & 65.0 \\
4sARP-NA & 71.3 & 28.8 & 161 & 76.6 & 310 & 147 & 520 & 195 & 100 & 7.54 & 191 & 11.6 & 360 & 14.1 & 630 & 41.0 & 114 & 10.4 & 190 & 25.3 & 383 & 69.6 & 619 & 55.2 \\
60sPre & 7.53 & 2.79 & 13.1 & 4.94 & 22.9 & 7.92 & 39.9 & 11.9 & 40.6 & 16.8 & 51.5 & 15.7 & 64.1 & 12.5 & 77.4 & 9.16 & 1.66 & .600 & 2.75 & 1.59 & 4.55 & 1.84 & 16.2 & 7.13 \\
60sBase\cite{Rasmussen2024} & 3.68 & .989 & 7.85 & 3.80 & 17.7 & 8.80 & 36.6 & 13.5 & 15.3 & 2.59 & 25.5 & 3.79 & 38.6 & 5.24 & 55.7 & 7.47 & 2.63 & 1.56 & 6.60 & 4.16 & 16.4 & 7.97 & 33.7 & 19.1 \\
60sDense & 2.00 & .542 & 3.52 & 1.02 & 7.17 & 1.89 & 16.3 & 3.70 & 5.31 & 1.88 & 9.05 & 3.28 & 15.7 & 5.08 & 28.6 & 8.84 & 3.01 & 2.85 & 4.93 & 4.26 & 9.29 & 6.04 & 21.9 & 14.7 \\
ARP-N & 1.17 & .018 & \textbf{1.61} & \textbf{.052} & \textbf{2.47} & .149 & \textbf{4.04} & \textbf{.190} & 1.94 & .185 & 3.09 & .510 & 5.31 & 1.06 & 11.1 & 2.79 & .864 & .127 & 1.03 & .245 & 1.27 & .335 & \textbf{1.71} & \textbf{.291} \\
ARPA-FN & 1.24 & .093 & 1.68 & .128 & \textbf{2.49} & \textbf{.144} & \textbf{4.13} & .345 & \textbf{1.83} & .099 & \textbf{2.55} & \textbf{.130} & \textbf{4.11} & \textbf{.271} & \textbf{7.93} & \textbf{.772} & .932 & .111 & 1.18 & .123 & 1.45 & .170 & 2.19 & .372 \\
ARPA-N & \textbf{1.11} & \textbf{.045} & \textbf{1.55} & .077 & \textbf{2.38} & .164 & 4.33 & .422 & \textbf{1.91} & \textbf{.071} & 2.97 & .209 & 5.47 & .938 & 12.9 & 3.15 & \textbf{.725} & \textbf{.056} & \textbf{.859} & \textbf{.111} & \textbf{1.11} & \textbf{.086} & \textbf{1.72} & .316 \\

\bottomrule
\end{tabularx}

\begin{tabularx}{1\textwidth}{|>{\columncolor{gray!15}}l|XXXXXXXXXl|XXXXXXXXXl|lcr|}
\toprule
\rowcolor{gray!35}
\multicolumn{1}{|c|}{} & \multicolumn{10}{|c|}{\textbf{Micro}} & \multicolumn{10}{c|}{\textbf{Macro}} & \multicolumn{3}{c|}{\textbf{Efficiency}} \\
\midrule
\rowcolor{gray!15}
\textbf{Model} & \textbf{AP} & $\sigma$ & \textbf{Rec} & $\sigma$ & \textbf{Pre} & $\sigma$ & \textbf{F1} & $\sigma$ & \textbf{R90} & $\sigma$ & \textbf{AP} & $\sigma$ & \textbf{Rec} & $\sigma$ & \textbf{Pre} & $\sigma$ & \textbf{F1} & $\sigma$ & \textbf{R90} & $\sigma$ & \textbf{ITA} & \textbf{ITS} & \textbf{Params} \\
\midrule

4sBase\cite{Miller2022} & .304 & .028 & .456 & .107 & .664 & .099 & .510 & .058 & 1020 & 63.7 & .218 & .023 & .447 & .136 & .490 & .097 & .418 & .056 & 1030 & 89.4 & 40.45 & .00015 & 6,897,021 \\
4sRes & .308 & .034 & .185 & .094 & .789 & .106 & .261 & .105 & 912 & 100 & .218 & .027 & .177 & .093 & \textbf{.646} & .175 & .236 & .102 & 946 & 129 & 36.12 & .00012 & 1.14e+7 \\
4sSCNN & .325 & .018 & .278 & .384 & .048 & .230 & .015 & .262 & 835 & 60.3 & .230 & .015 & .262 & .054 & .549 & .041 & .332 & .049 & 889 & 74.9 & 28.46 & .0001 & 9,915,457 \\
4sARP-N & .353 & .018 & .634 & .663 & .586 & .057 & .572 & .047 & 891 & 74.8 & .256 & .017 & .609 & .057 & .427 & .054 & .468 & .049 & 925 & 62.8 & 24.79 & 8e-5 & 1,167,681 \\
4sResA & .436 & .062 & .568 & .078 & .335 & .041 & .564 & .087 & 785 & 116 & .335 & .041 & .564 & .087 & .497 & .041 & .492 & .036 & 696 & 92.9 & 36.12 & .00012 & 1.14e+7 \\
4sARP-NA & .507 & .017 & .670 & .777 & .556 & .057 & .570 & .027 & 557 & 143 & .379 & .016 & .644 & .081 & .387 & .044 & .443 & .025 & 589 & 97.0 & 24.79 & 8e-5 & 1,167,681 \\
60sPre & .560 & .074 & .690 & .078 & .645 & .066 & .422 & .079 & 50.9 & 10.9 & .422 & .079 & .709 & .087 & .438 & .085 & .489 & .082 & 44.5 & 9.39 & 37.84 & .00172 & 2.41e+7 \\
60sBase\cite{Rasmussen2024} & .665 & .017 & .742 & .010 & .527 & .019 & .724 & .025 & 42.5 & 11.8 & .527 & .019 & .724 & .025 & .583 & .029 & .608 & .015 & 42.0 & 13.3 & 12.18 & .00055 & 1.22e+7 \\
60sDense & .727 & .024 & .804 & .025 & .781 & .015 & .781 & .018 & 20.2 & 5.59 & .572 & .018 & .752 & .033 & .628 & .042 & .673 & .027 & 22.2 & 9.07 & 40.45 & .00183 & 6,897,021 \\
ARP-N & .793 & .006 & \textbf{.866} & .024 & .734 & .027 & .790 & .007 & 6.18 & 1.00 & .630 & .014 & \textbf{.838} & .026 & .559 & .031 & .639 & .016 & 5.63 & 1.09 & 17.19 & .00078 & 4,968,769 \\
ARPA-FN & .798 & .007 & \textbf{.843} & \textbf{.022} & .762 & .020 & .796 & .007 & \textbf{5.26} & \textbf{.478} & .628 & .009 & \textbf{.826} & \textbf{.020} & .592 & .018 & .658 & .009 & \textbf{4.75} & \textbf{.496} & 21.59 & .00098 & 4,968,872 \\
ARPA-N & \textbf{.809} & \textbf{.008} & .788 & .022 & \textbf{.831} & \textbf{.022} & \textbf{.806} & \textbf{.004} & 6.92 & 1.27 & \textbf{.656} & \textbf{.011} & .750 & .029 & \textbf{.664} & \textbf{.020} & \textbf{.687} & \textbf{.010} & 6.32 & 1.30 & 27.78 & .00126 & 4,969,387 \\

\bottomrule
\end{tabularx}
\caption{Comparative Results. This table reports AP, Recall, Precision, F1, and FP/hr (with variance $\sigma$) for Datasets~1–11, focusing on Kerguelen~2015, Casey~2017, and BallenyIslands~2015, which span different annotation levels and yield micro/macro metrics. The final section lists inference time (IT, s) for BallenyIslands~2015 and a single sonograph, along with parameter count.}
\label{tab:res}
\end{table*}

\subsection{Baseline Architectures}

Table~\ref{tab:res} reports AP, F1, Precision, Recall, FP/hr, and Micro/Macro aggregates for all 12 baseline configurations under GetNetUPAM across Kerguelen~2015 (2‑year support), Casey~2017 (1‑year support), and Balleny~2015 (zero support). Models are ordered by Micro AP, which treats all positives equally under the binary‑relevance formulation and provides a single threshold‑independent summary of the precision–recall trade‑off. We report AP rather than full PR curves, as AP is the standard AUC‑equivalent summary.

\paragraph{Short‑window baselines (4s)} The DenseNet baseline \cite{Miller2022} (4sBase) sets the lower bound for 4‑s preprocessing (Kerguelen F1 = .575, Macro F1 = .418, Macro R90 = 1030 FP/hr). The feedforward CNN (4sSCNN) increases precision (.877) but collapses recall (.241), yielding poor F1 (.375). 4sARP‑N recovers moderate performance (Kerguelen F1 = .666) and reduces Macro R90 to 925 FP/hr, isolating the benefit of adaptive pooling. Utilizing all of the dataset (4sResA, 4sARP‑NA) improve Casey F1 (.553 and .385) and reduce Macro R90 (696 and 589 FP/hr) but remain far below all 60‑s models, confirming window length as the dominant factor for D‑call capture. However, curated variants that discard clipped positives and negatives perform worse than non‑curated versions. Retaining truncated positives improves Casey AP by 0.20 and 0.17 and Balleny AP by 0.053 and 0.051, showing that removing too many samples systematically harms performance.

\paragraph{Long‑window baselines (60s)} The pretrained ResNet‑50 (60sPre) achieves strong Kerguelen F1 (.759) but shows high instability at Casey (R90 $\sigma$ = 9.16) and weak zero‑support transfer (Balleny R90 = 16.2 FP/hr). The ResNet‑18 model from Rasmussen~\cite{Rasmussen2024} (\textbf{60sBase}) provides a more stable baseline, with Kerguelen F1 = .799 ($\sigma$ = .006), Casey F1 = .650, and Macro R90 = 42.0 FP/hr, though Balleny precision remains low (.248). DenseNet‑60s is the strongest standard architecture long‑window baseline, achieving Kerguelen F1 = .837, Casey F1 = .687, Macro AP = .572, and Macro R90 = 22.2 FP/hr, establishing the upper bound for comparison with ARPA‑N and ARPA‑FN.

\paragraph{ViTs and Foundation Models} Vision Transformers and cross‑taxa foundation models were evaluated but excluded from Table~\ref{tab:res} due to consistent failure to learn the D‑call task. All ViTs \cite{dosovitskiy2021vit} and variants converged to near‑zero recall with undefined precision, indicating incompatibility between patch‑based tokenization/global self‑attention and the sparse, low‑SNR structure of D‑calls. Cross‑taxa transfer with SurfPerch \cite{williams2025using} also degraded severely due to the mismatch between our 250\,Hz target rate and its 32\,kHz native representation. Because these models did not reach a minimally functional operating point, quantitative comparison would be misleading; their exclusion reflects task–model mismatch rather than selective reporting.

\begin{table*}[t!]
\scriptsize
\centering
\begin{tabularx}{\textwidth}{|>{\columncolor{gray!15}}l|XXXXXXXl|XXXXXXXl|XXXXXXXl|}
\toprule
\rowcolor{gray!35}
\multicolumn{1}{|c}{} & \multicolumn{4}{|c}{\textbf{Kerguelen 2015}} & \multicolumn{4}{c|}{\textbf{2 Years Support}} & \multicolumn{4}{c}{\textbf{Casey 2017}} & \multicolumn{4}{c|}{\textbf{1 Year Support}} & \multicolumn{4}{c}{\textbf{BallenyIslands 2015}} & \multicolumn{4}{c|}{\textbf{No Support}} \\
\midrule
\rowcolor{gray!15}
\textbf{Model} & \textbf{AP} & $\sigma$ & \textbf{Rec} & $\sigma$ & \textbf{Pre} & $\sigma$ & \textbf{F1} & $\sigma$ & \textbf{AP} & $\sigma$ & \textbf{Rec} & $\sigma$ & \textbf{Pre} & $\sigma$ & \textbf{F1} & $\sigma$ & \textbf{AP} & $\sigma$ & \textbf{Rec} & $\sigma$ & \textbf{Pre} & $\sigma$ & \textbf{F1} & $\sigma$ \\
\midrule

Rand Flip & .625 & .201 & .604 & .227 & .864 & .020 & .685 & .184 & .613 & .032 & .640 & .042 & .734 & .016 & .682 & .020 & .330 & .028 & .719 & .065 & .323 & .067 & .438 & .054 \\
Vanilla & .776 & .010 & .661 & .015 & \textbf{.959} & \textbf{.011} & .782 & .008 & .619 & .007 & .544 & .035 & \textbf{.847} & \textbf{.036} & .661 & .013 & .310 & .018 & .477 & .063 & \textbf{.455} & \textbf{.057} & \textbf{.461} & .040 \\
S+D+K+B & .800 & .053 & .830 & .023 & .853 & .030 & .841 & .006 & .676 & .011 & .713 & .009 & .723 & .009 & .718 & .005 & \textbf{.348} & .038 & .728 & .021 & .330 & .024 & \textbf{.453} & \textbf{.019} \\
All - K & .824 & .006 & .872 & .036 & .794 & .051 & .829 & .012 & .677 & .005 & \textbf{.810} & \textbf{.022} & .568 & .039 & .666 & .020 & .307 & .029 & .800 & .074 & .261 & .042 & .389 & .035 \\
All - D & .831 & .006 & .883 & .016 & .802 & .024 & .840 & .007 & .683 & .011 & .763 & .037 & .628 & .046 & .687 & .011 & .332 & .032 & .809 & .036 & .251 & .018 & .382 & .022 \\
All - G & .830 & .007 & \textbf{.893} & .025 & .785 & .033 & .834 & .008 & .687 & .019 & \textbf{.809} & .023 & .592 & .017 & .683 & .009 & .324 & .021 & \textbf{.826} & .045 & .240 & .023 & .372 & .030 \\
All - S & .812 & .006 & .844 & .016 & .813 & .018 & .828 & .009 & .725 & .004 & .778 & .042 & .667 & .027 & .716 & .010 & .335 & .028 & \textbf{.864} & .044 & .240 & .017 & .375 & .023 \\
CANN & .830 & .009 & .792 & .039 & .859 & .023 & .823 & .011 & .718 & .019 & .680 & .033 & .741 & .028 & .708 & .013 & \textbf{.371} & .013 & .817 & .032 & .244 & .022 & .376 & .027 \\
All + CAF & .836 & .009 & .878 & .027 & .796 & .034 & .834 & .009 & .733 & .011 & \textbf{.796} & .031 & .649 & .026 & .714 & .010 & \textbf{.347} & .025 & \textbf{.821} & .044 & .251 & .025 & .384 & .031 \\
CRNN & .839 & .005 & .846 & .018 & .840 & .030 & .838 & .007 & .727 & .011 & .745 & .027 & .686 & .029 & .713 & .011 & \textbf{.354} & .028 & .813 & .025 & .234 & .014 & .363 & .019 \\
All & \textbf{.850} & .005 & \textbf{.903} & \textbf{.017} & .794 & .019 & .845 & .004 & .710 & .005 & \textbf{.790} & .039 & .648 & .044 & .709 & .011 & .330 & .030 & \textbf{.821} & .022 & .235 & .028 & .364 & .033 \\
All + CA & .843 & .006 & \textbf{.907} & .017 & .763 & .036 & .828 & .015 & .724 & .021 & \textbf{.823} & .041 & .583 & .041 & .680 & .019 & \textbf{.365} & .021 & \textbf{.834} & .045 & .233 & .017 & .363 & .019 \\
All + MPF & .847 & .010 & .887 & .015 & .806 & .017 & .844 & .004 & .724 & .008 & .780 & .033 & .663 & .041 & .715 & .013 & .328 & .044 & .817 & .029 & .247 & .040 & .377 & .045 \\
All - M & \textbf{.855} & .008 & \textbf{.891} & .017 & .817 & .019 & \textbf{.852} & .006 & .711 & .017 & .779 & .039 & .665 & .030 & .716 & .010 & \textbf{.354} & \textbf{.012} & \textbf{.843} & .029 & .241 & .020 & .374 & .023 \\
All + SAF & \textbf{.852} & .006 & .883 & .023 & .812 & .021 & .845 & .007 & .722 & .009 & .756 & .018 & .697 & .019 & .725 & .007 & .309 & .011 & \textbf{.838} & \textbf{.017} & .267 & .014 & .404 & .014 \\
All + SA & \textbf{.857} & .008 & .823 & .023 & .888 & .023 & \textbf{.854} & \textbf{.003} & \textbf{.743} & \textbf{.007} & .720 & .019 & .749 & .020 & \textbf{.734} & \textbf{.003} & \textbf{.367} & .017 & .706 & .045 & .357 & .019 & \textbf{.474} & .023 \\
All + MP & \textbf{.856} & \textbf{.004} & .855 & .040 & .841 & .036 & .846 & .003 & \textbf{.747} & .009 & .730 & .016 & .721 & .035 & .725 & .014 & \textbf{.372} & .027 & .783 & .072 & .313 & .031 & .444 & .021 \\

\bottomrule
\end{tabularx}
\begin{tabularx}{\textwidth}{|>{\columncolor{gray!15}}l|XXXXXXXl|XXXXXXXl|XXXXXXXl|}
\toprule
\rowcolor{gray!15}
\textbf{Model} & \textbf{R75} & $\sigma$ & \textbf{R80} & $\sigma$ & \textbf{R85} & $\sigma$ & \textbf{R90} & $\sigma$ & \textbf{R75} & $\sigma$ & \textbf{R80} & $\sigma$ & \textbf{R85} & $\sigma$ & \textbf{R90} & $\sigma$ & \textbf{R75} & $\sigma$ & \textbf{R80} & $\sigma$ & \textbf{R85} & $\sigma$ & \textbf{R90} & $\sigma$ \\ 
\midrule

Rand Flip & 12.3 & 13.1 & 15.9 & 15.2 & 22.0 & 16.9 & 32.0 & 17.1 & 8.20 & 3.78 & 13.6 & 5.81 & 21.9 & 7.04 & 33.3 & 7.56 & 1.95 & .742 & 4.18 & 3.71 & 6.88 & 4.41 & 19.5 & 13.1 \\
Vanilla & 2.64 & .539 & 5.20 & .596 & 9.52 & .941 & 18.8 & 3.21 & 13.2 & 1.56 & 21.8 & 1.47 & 34.1 & .915 & 49.0 & 1.89 & 2.66 & .813 & 3.61 & .940 & 6.12 & 1.96 & 14.0 & 6.98 \\
S+D+K+B & 1.60 & .154 & 2.67 & .338 & 5.89 & 1.22 & 16.9 & 3.98 & 3.06 & .506 & 5.87 & .793 & 13.1 & 1.46 & 25.7 & 1.51 & 1.21 & .171 & 2.03 & .322 & 6.08 & 6.27 & 14.6 & 7.54 \\
All - K & 1.48 & .105 & 2.09 & .075 & 3.22 & .248 & 5.77 & .465 & 2.73 & .132 & 4.16 & .301 & 6.67 & .369 & 13.0 & 1.43 & 1.01 & .260 & 1.39 & .522 & 1.76 & .710 & 4.04 & 3.18 \\
All - D & 1.33 & .046 & 1.95 & .035 & 3.16 & .266 & 5.98 & .901 & 2.78 & .247 & 4.18 & .432 & 7.31 & .610 & 14.8 & 1.49 & .849 & .046 & .989 & .093 & 1.25 & .130 & 2.88 & 1.63 \\
All - G & 1.37 & .045 & 1.99 & .147 & 3.06 & .228 & 5.21 & .507 & 2.30 & .266 & 3.53 & .539 & 6.25 & 1.10 & 12.6 & 2.55 & .858 & .081 & 1.17 & .082 & 1.65 & .173 & 2.93 & .839 \\
All - S & 1.56 & .112 & 2.07 & .173 & 3.18 & .217 & 5.14 & .370 & 1.96 & .145 & 3.06 & .177 & 5.06 & .250 & 10.6 & .811 & .788 & .184 & .901 & .197 & 1.04 & .224 & 1.59 & .353 \\
CANN & 1.57 & .143 & 2.17 & .163 & 2.96 & .252 & 4.95 & .369 & 2.43 & .654 & 3.84 & .955 & 7.13 & 2.15 & 14.7 & 3.89 & \textbf{.687} & .064 & \textbf{.777} & .102 & \textbf{.940} & .110 & \textbf{1.31} & .176 \\
All + CAF & 1.40 & .104 & 1.94 & .153 & 2.78 & .218 & 4.78 & .430 & \textbf{1.89} & .217 & \textbf{2.67} & .369 & \textbf{4.34} & .644 & \textbf{8.66} & .873 & .826 & .180 & .951 & .152 & 1.22 & .175 & 1.64 & .283 \\
CRNN & 1.36 & .109 & 1.91 & .125 & 2.94 & .282 & 5.08 & .659 & \textbf{1.86} & .127 & 2.87 & .266 & 5.07 & .727 & 10.0 & 1.62 & .752 & .032 & \textbf{.871} & \textbf{.022} & 1.07 & .114 & \textbf{1.39} & \textbf{.174} \\
All & 1.17 & .018 & \textbf{1.61} & \textbf{.053} & \textbf{2.47} & .149 & \textbf{4.04} & .190 & 1.94 & .185 & 3.09 & .510 & 5.31 & 1.06 & 11.1 & 2.79 & .864 & .127 & 1.03 & .245 & 1.27 & .335 & 1.71 & .291 \\
All + CA & 1.34 & .097 & 1.86 & .131 & 2.65 & .175 & \textbf{4.33} & .250 & 2.10 & .395 & 3.03 & .711 & 4.85 & 1.45 & \textbf{8.45} & 2.53 & .774 & .060 & .975 & .183 & 1.1 & .182 & 1.55 & .299 \\
All + MPF & 1.27 & .057 & 1.77 & .138 & \textbf{2.53} & .115 & \textbf{4.28} & \textbf{.160} & \textbf{1.87} & .106 & \textbf{2.66} & .198 & \textbf{4.26} & \textbf{.106} & \textbf{8.47} & \textbf{.611} & .980 & .114 & 1.09 & .125 & 1.38 & .265 & 2.12 & .560 \\
All - M & \textbf{1.14} & .086 & \textbf{1.56} & .102 & \textbf{2.37} & .244 & \textbf{3.93} & .450 & 1.95 & .322 & 3.23 & .899 & 6.29 & 1.68 & 13.3 & 2.81 & \textbf{.677} & \textbf{.040} & \textbf{.826} & .079 & \textbf{.936} & \textbf{.089} & \textbf{1.29} & .177 \\
All + SAF & 1.24 & .093 & 1.68 & .128 & \textbf{2.49} & .144 & \textbf{4.13} & .345 & \textbf{1.83} & .099 & \textbf{2.55} & \textbf{.130} & \textbf{4.11} & .271 & \textbf{7.93} & .772 & .932 & .111 & 1.18 & .123 & 1.45 & .170 & 2.19 & .372 \\
All + SA & \textbf{1.11} & \textbf{.045} & \textbf{1.55} & .077 & \textbf{2.38} & .164 & \textbf{4.33} & .422 & \textbf{1.91} & \textbf{.071} & 2.97 & .209 & 5.47 & .938 & 12.9 & 3.15 & .725 & .057 & \textbf{.859} & .111 & 1.11 & .086 & 1.72 & .316 \\
All + MP & 1.17 & .088 & \textbf{1.60} & .106 & \textbf{2.37} & \textbf{.097} & \textbf{4.03} & .391 & \textbf{1.83} & .212 & 2.75 & .357 & \textbf{4.36} & .504 & 8.91 & .620 & .731 & .127 & \textbf{.848} & .108 & 1.04 & .130 & 1.50 & .427 \\

\bottomrule
\end{tabularx}
\begin{tabularx}{1\textwidth}{|>{\columncolor{gray!15}}l|XXXXXXXXXl|XXXXXXXXXl|lcr|}
\toprule
\rowcolor{gray!35}
\multicolumn{1}{|c|}{} & \multicolumn{10}{|c|}{\textbf{Micro}} & \multicolumn{10}{c|}{\textbf{Macro}} & \multicolumn{3}{c|}{\textbf{Efficiency}} \\
\midrule
\rowcolor{gray!15}
\textbf{Model} & \textbf{AP} & $\sigma$ & \textbf{Rec} & $\sigma$ & \textbf{Pre} & $\sigma$ & \textbf{F1} & $\sigma$ & \textbf{R90} & $\sigma$ & \textbf{AP} & $\sigma$ & \textbf{Rec} & $\sigma$ & \textbf{Pre} & $\sigma$ & \textbf{F1} & $\sigma$ & \textbf{R90} & $\sigma$ & \textbf{ITA} & \textbf{ITS} & \textbf{Params} \\
\midrule

Rand Flip & .613 & .144 & .618 & .165 & .809 & .020 & .677 & .130 & 32.0 & 14.0 & .523 & .087 & .654 & .111 & .640 & .034 & .602 & .086 & 28.2 & 12.6 & 19.47 & .00088 & 9,163,073 \\
Vanilla & .715 & .009 & .620 & .022 & \textbf{.911} & \textbf{.020} & .736 & .011 & 28.0 & 2.90 & .569 & .011 & .561 & .038 & \textbf{.754} & \textbf{.034} & .635 & .021 & 27.3 & 4.03 & 12.64 & .00057 & 9,057,793 \\
S+D+K+B & .749 & .008 & .791 & .018 & .799 & .023 & .792 & .006 & 19.6 & 3.31 & .608 & .018 & .757 & .018 & .635 & .021 & .671 & .018 & 19.1 & 4.34 & 19.47 & .00088 & 9,163,073 \\
All - K & .765 & .006 & .851 & .033 & .710 & .047 & .767 & .015 & 7.97 & .838 & .603 & .013 & \textbf{.827} & .044 & .541 & .044 & .628 & .022 & 7.60 & 1.69 & 14.22 & .00065 & 4,900,673 \\
All - D & .772 & .008 & .843 & .023 & .733 & .031 & .780 & .008 & 8.65 & 1.10 & .615 & .018 & .818 & .030 & .560 & .029 & .636 & .013 & 7.90 & 1.34 & 9.17 & .00042 & 4,866,049 \\
All - G & .772 & .011 & \textbf{.855} & .025 & .710 & .028 & .775 & .009 & 7.45 & 1.15 & .614 & .015 & \textbf{.842} & .031 & .635 & .036 & .630 & .016 & 6.92 & 1.30 & 17.21 & .00078 & 4,968,769 \\
All - S & .772 & .006 & .824 & .025 & .752 & .021 & .781 & .009 & 6.75 & .506 & .624 & .013 & \textbf{.828} & .034 & .573 & .021 & .640 & .014 & 5.78 & .511 & 17.2 & .00078 & 4,968,769 \\
CANN & .783 & .013 & .758 & .037 & .806 & .024 & .775 & .012 & 7.89 & 1.46 & .640 & .014 & .763 & .035 & .615 & .024 & .636 & .017 & 6.99 & 1.48 & 17.5 & .00079 & 2,218,817 \\
All + CAF & .791 & .010 & .851 & .029 & .736 & .031 & .785 & .010 & 5.90 & .564 & .639 & .015 & \textbf{.832} & .034 & .565 & .028 & .644 & .017 & \textbf{5.02} & .528 & 21.51 & .00096 & 4,985,441 \\
CRNN & .791 & .008 & .814 & .021 & .770 & .022 & .786 & .008 & 6.51 & .944 & .640 & .015 & .801 & .023 & .583 & .021 & .638 & .012 & 5.50 & .817 & 20.45 & .00093 & 2,217,793 \\
All & .793 & .006 & \textbf{.866} & \textbf{.024} & .734 & .027 & .790 & .007 & 6.18 & .999 & .630 & .014 & \textbf{.838} & .026 & .559 & .031 & .639 & .016 & 5.63 & 1.09 & 17.19 & .00078 & 4,968,769 \\
All + CA & .794 & .011 & \textbf{.879} & .025 & .693 & .037 & .770 & .016 & \textbf{5.53} & .960 & .644 & .016 & \textbf{.855} & .035 & .526 & .031 & .624 & .017 & \textbf{4.78} & 1.03 & 25.72 & .00116 & 5,000,353 \\
All + MPF & .795 & .008 & .852 & .021 & .747 & .025 & .792 & .008 & \textbf{5.52} & \textbf{.310} & .633 & .020 & \textbf{.828} & .026 & .572 & .033 & .645 & .021 & \textbf{4.95} & \textbf{.443} & 21.64 & .00098 & 4,985,544 \\
All - M & .797 & .011 & \textbf{.855} & .024 & .755 & .023 & .797 & .008 & 6.76 & 1.18 & .640 & .013 & \textbf{.837} & .028 & .574 & .023 & .647 & .013 & 6.16 & 1.15 & 19.49 & .00088 & 9,163,073 \\
All + SAF & .798 & .007 & .843 & .022 & .762 & .020 & .796 & .007 & \textbf{5.26} & .478 & .628 & .009 & \textbf{.826} & \textbf{.020} & .592 & .018 & .658 & .009 & \textbf{4.75} & .496 & 21.59 & .00098 & 4,968,872 \\
All + SA & \textbf{.809} & .008 & .788 & .022 & .831 & .022 & \textbf{.806} & \textbf{.004} & 6.92 & 1.27 & \textbf{.656} & \textbf{.011} & .750 & .029 & .664 & .020 & \textbf{.687} & \textbf{.010} & 6.32 & 1.30 & 27.78 & .00126 & 4,969,387 \\
All + MP & \textbf{.810} & \textbf{.006} & .814 & .033 & .790 & .035 & .798 & .007 & \textbf{5.48} & .463 & \textbf{.659} & .013 & .789 & .043 & .625 & .034 & .672 & .013 & \textbf{4.81} & .479 & 34.2 & .00155 & 5,000,971 \\

\bottomrule
\end{tabularx}
\caption{Ablation of Network Components. This table reports AP, Recall, Precision, F1, and FP/hr (with variance $\sigma$) for Datasets~1–11, focusing on Kerguelen~2015, Casey~2017, and BallenyIslands~2015, which span different annotation levels and yield micro/macro metrics. The final section lists inference time (IT, s) for BallenyIslands~2015 and a single sonograph, along with parameter count.}
\label{tab:ab}
\end{table*}

\subsection{Results and Comparative Analysis}

\paragraph{Kerguelen 2015 (2-Year Support).} ARP‑N performs strongly on this best‑supported site (AP = .850, F1 = .845), while ARPA‑N yields the top overall results (AP = .857, Precision = .888, F1 = .854) with the tightest fold‑level F1 ($\sigma$ = .003). Relative to DenseNet‑60s, ARPA‑N improves AP by 6.6 points, Precision by 6.9, and F1 by 1.7, while reducing F1 variance 3.6× (.011 → .003). ARP‑N matches ARPA‑FN on F1 (.845), but provides the best Recall and R90 metrics.

\paragraph{Casey 2017 (1-Year Support).} ARPA‑N leads on AP (.744), Precision (.756), and F1 (.733), outperforming DenseNet‑60s by 11.7 AP points, 7.8 Precision points, and 4.6 F1 points. ARP‑N attains the highest Recall (.790) but with lower Precision (.648) and F1 (.709). ARPA‑FN offers the most stable operational profile, with the lowest Recall variance ($\sigma$ = .018) and the lowest R90 (7.93 FP/hr) among ARP variants.

\paragraph{Balleny Islands 2015 (Zero Support).} DenseNet‑60s achieves the highest raw F1 (.495, $\sigma$ = .052) but with extremely high variance and FP/hr (R90 = 21.9 FP/hr, $\sigma$ = 14.7). ARPA‑N attains F1 = .474 with AP = .367 and R90 = 1.72 FP/hr—a 12.7× reduction in false positives at comparable F1. ARP‑N yields the lowest R90 (1.71 FP/hr) but with substantially lower F1 (.364). ARPA‑FN provides the tightest F1 estimate ($\sigma$ = .014) with F1 = .404 and R90 = 2.19 FP/hr.

\paragraph{Aggregate: Micro and Macro.} ARPA‑N leads all models on Micro AP (.809), Precision (.831), F1 (.806), Macro AP (.656), Precision (.664), and F1 (.687) corresponding to gains of 8.2, 5.0, 2.5, 8.4, 3.6, and 1.5 points over DenseNet‑60s. ARP‑N leads on Micro Recall (.866) and Macro Recall (.838), reflecting its recall‑oriented behavior. ARPA‑FN achieves the lowest Micro and Macro R90 (5.26 and 4.75 FP/hr), reducing FP/hr 3.8× and 4.7× and an order of magnitude drop in variance relative to DenseNet‑60s. ARPA‑N’s Macro AP improvement (.656 vs.\ .572) produces a 14.7\% relative gain.

\paragraph{Efficiency.} ARPA‑N runs at 0.00126\,s per spectrogram (27.78\,s for Balleny) with 4.97M parameters—smaller than DenseNet‑60s (6.90M) and far smaller than ResNet‑50 (24M) or 60sBase (12.2M)—while outperforming all three on IR metrics. ARPA‑FN matches ARP‑N in parameter count and inference time (21.6\,ms vs.\ 17.2\,ms) and is preferred when minimizing FP/hr is the priority. All$-$D offers an edge‑deployment option at 4.87M parameters and 9.17\,ms inference while retaining Kerguelen F1 within 0.5 points of ARPA‑N.

\begin{figure*}[t!] \centering \includegraphics[width=1.0\linewidth]{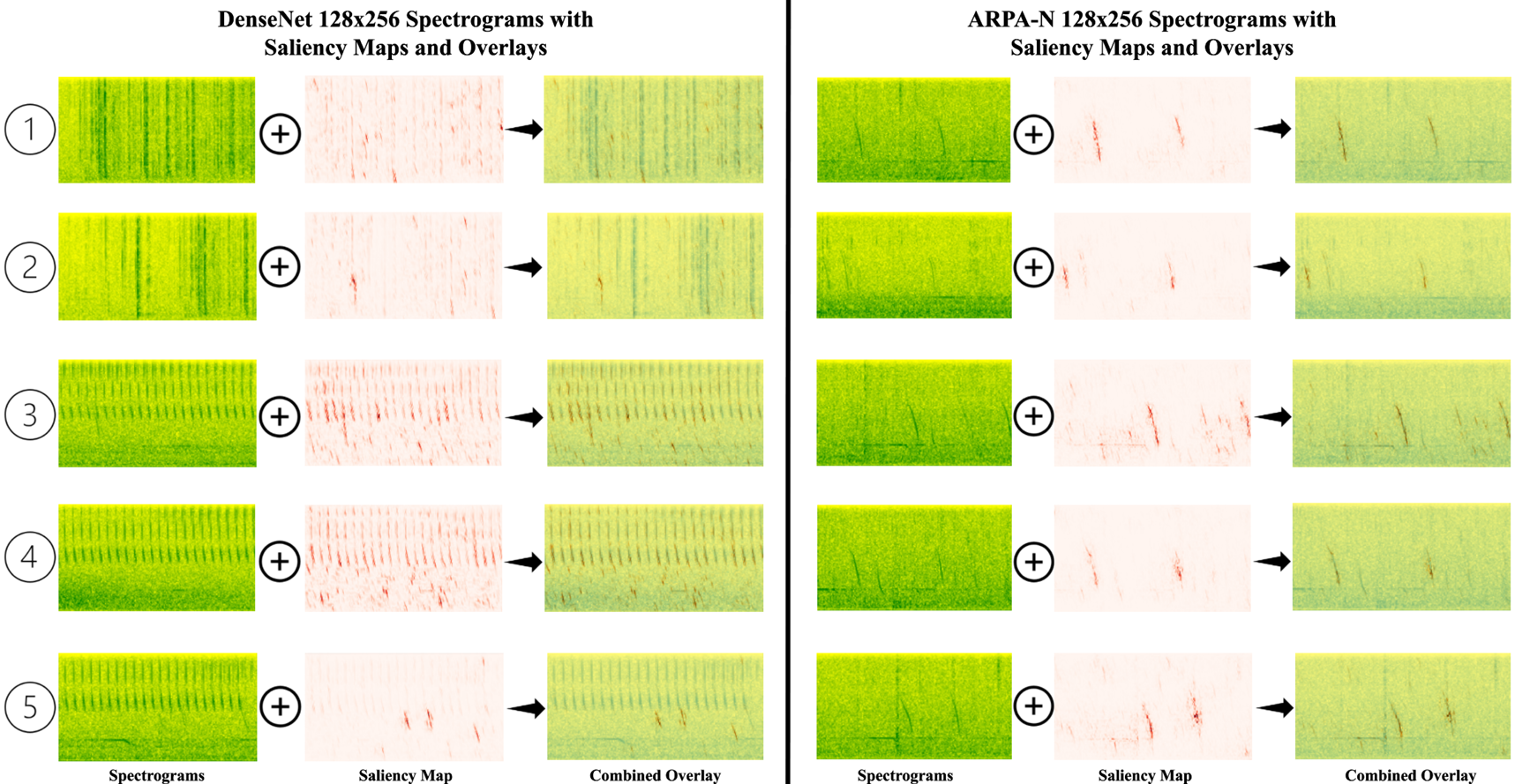} \caption{Comparative Analysis of Acoustic Event Detection. The figure illustrates saliency map overlays on human-interpretable spectrograms for whale call detection. On the left, saliency maps from DenseNet display sporadic patterns, with only a subset aligning with whale calls. ARPA-N saliency maps on the right reveal pronounced bright areas that accurately coincide with whale D-Calls, demonstrating enhanced temporal localization. This contrast underscores the efficacy of ARPA-N in facilitating precise event detection and enabling robust human-AI collaboration for UPAM data exploration.}\label{fig:soDiag} \end{figure*}

\subsection{Ablation Study: Network Components}

Table~\ref{tab:ab} reports AP, Recall, Precision, F1, and FP/hr at four thresholds (R75–R90) for 17 ablation configurations across three site–years (Kerguelen~2015, Casey~2017, Balleny~2015), along with Micro/Macro aggregates and efficiency metrics. All values include fold‑level $\sigma$ from the nested CV. Our model was built incrementally using the following components:
\textbf{S}: sigmoid output; 
\textbf{D}: second initial convolution; 
\textbf{K}: early kernel-size adjustment; 
\textbf{B}: Initial spatial dropout; 
\textbf{G}: Gaussian noise; 
\textbf{M}: adaptive pooling; 
\textbf{All}: + spatial dropout throughout; 
\textbf{CANN}: post‑CNN multi‑head attention; 
\textbf{CRNN}: post‑CNN recurrent block; 
\textbf{CA}: CBAM channel attention; 
\textbf{SA}: CBAM spatial attention; 
\textbf{MP}: multi‑path CA+SA; 
\textbf{F}: final layer attention.

\paragraph{Component Contributions.}
Horizontal flip augmentation produces the highest instability, confirming orientation sensitivity in D‑calls. The Vanilla model attains high precision on the best‑supported site but fails cross‑site, showing that unregularized CNNs overfit spectral coloration. The S+D+K+B combination restores cross‑site performance and reduces false positives, forming the minimal stable backbone. Kernel‑size adjustment (\textbf{K}) and sigmoid output (\textbf{S}) contribute most to stability; removing the second convolution (\textbf{D}) yields the most efficient configuration still within 0.5 F1 of the full model.

\paragraph{Alternative Sequence Architectures.}
Post‑CNN transformer attention (CANN) and CRNNs reduce parameters but do not improve cross‑site generalization; both underperform on the zero‑support site, indicating that long‑range temporal modeling alone does not address ATBFL noise‑structure variability. Thus, ARPA‑N’s gains arise from spatially targeted feature selection, not sequence modeling.

\paragraph{Attention Variants.}
Full CBAM spatial attention (\textbf{All+SA}) provides the strongest IR performance, improving Macro F1, AP, and Precision with the tightest F1 stability. Multi‑path attention (\textbf{All+MP}) slightly increases AP but reduces F1 and Precision. Channel‑only attention (\textbf{All+CA}) consistently underperforms, confirming spatial localization, not channel reweighting, as the key driver of D‑call discriminability.

\paragraph{False Positives Per Hour.}
Final‑layer spatial attention (\textbf{All+SAF}) yields the lowest Macro R90, while full‑depth SA trades higher FP/hr for stronger IR metrics. This reflects a threshold‑dependent tradeoff: full‑depth SA improves discriminative power; final‑layer SA suppresses late false activations.

\paragraph{Cross‑Site Generalization.}
\textbf{All+SA} shows the smallest Kerguelen→Balleny F1 drop and the highest Balleny F1, indicating strongest suppression of site‑specific shortcuts. \textbf{All+SAF} and \textbf{All+MP} generalize moderately; the ARP‑N base shows the largest degradation.

\paragraph{Inference Efficiency.}
\textbf{All$-$D} is the fastest configuration that maintains competitive F1. CANN and CRNN are parameter‑efficient but do not match the FP/hr performance of spatial‑attention variants. Multi‑path attention (\textbf{All+MP}) incurs the highest inference cost without commensurate gains, reinforcing full spatial attention as the best detection and stability tradeoff reinforcing our choice in the \textbf{All+SA model.}

\subsection{Human-Interpretable Saliency Mapping}

We ranked Casey~2017 samples by model probability and compared saliency maps from the DenseNet baseline \cite{Miller2022} (adapted to our preprocessing) with those from ARPA‑N. From each model, the five most informative examples (Fig.~\ref{fig:soDiag}) illustrate qualitative differences in attention behavior. DenseNet produces scattered, inconsistent activations: only one of the five samples aligns cleanly with D‑call contours, while the others highlight unrelated spectrogram regions, reflecting the difficulty short‑window baselines have in focusing on the target signal. ARPA‑N, in contrast, concentrates sharply on spectral–temporal regions matching D‑call signatures, providing consistent and anatomically plausible localization across all examples. Its heatmaps follow call trajectories closely, enabling precise temporal pinpointing and supporting automated event‑to‑timestamp mapping for downstream tabular or GIS workflows.

\begin{figure}[t!]
\centering
\includegraphics[width=1.0\linewidth]{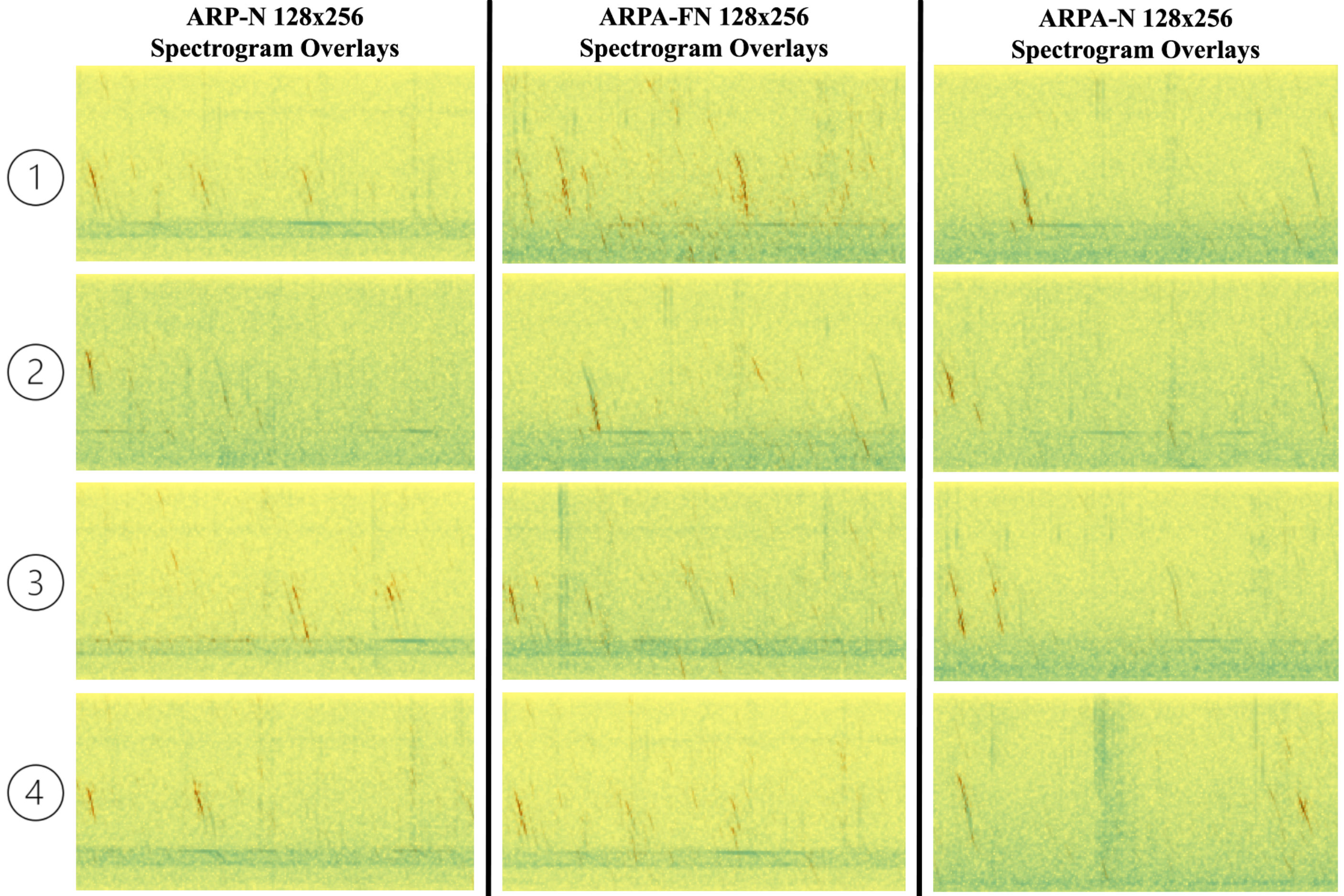}
\caption{Ablation of Acoustic Event Detection. Saliency map overlays on human-interpretable spectrograms for whale call detection. Left: ARP-N (ALL) maps display attention to mostly only Biological Pattern, with attention to non-D Blue Whale Calls. Middle: ARPA-FN (All+SAF) displays attention to only downswept features, though more sporadic. Right: ARPA-N (All+SA) saliency maps reveal sparser activations only targeting whale D-call showing difference in localization with spatial attention applied differently across layers.}
\label{fig:attnDiag}
\end{figure}

Figure~\ref{fig:attnDiag} further compares saliency behavior across ARP‑N, ARPA‑FN, and ARPA‑N, clarifying their inductive biases. ARP‑N generally attends to D‑calls but also activates on other Blue Whale vocalizations—most notably A‑type “anthem” calls—indicating residual sensitivity to broad spectral cues. ARPA‑FN focuses consistently on downswept D‑call energy but often produces saturated maps with limited selectivity for other features. ARPA‑N provides the cleanest and most selective localization: attention is sharply confined to D‑call contours with minimal spillover, yielding the most interpretable and biologically aligned saliency among the three architectures.

Beyond accuracy, ARPA‑N offers a clear interpretability gain: its localized overlays are more readable to human analysts than earlier DenseNet‑based visualizations, improving error‑checking in human‑in‑the‑loop pipelines and increasing confidence when triaging UPAM datasets for misclassifications \cite{Nunnari2021}. By making model decisions visually legible and target events easy to verify, ARPA‑N better aligns automated detection with expert ecological judgment.

\subsection{Operational Deployment Considerations.}
The operational analysis indicates that attention configuration should be selected according to the dominant deployment constraint, balancing discriminative performance, false–positive burden, and computational efficiency. The full–depth spatial–attention model (All+SA) provides the strongest overall detection capability—achieving the highest Macro~F1 (.687), Macro Precision (.664), and zero–support generalization—consistent with the observation that “CBAM spatial attention alters not only \emph{how well} ARPA–N learns but fundamentally \emph{what} it learns.” Its computational profile (4.97\,M parameters; 27.8\,ms inference) is suitable for server–class processing, though its elevated Casey R90 (12.9\,FP/hr) increases analyst load in continuous–monitoring scenarios. The final–layer spatial–attention variant (All+SAF) shifts the operating point toward operational stability, reducing Macro~R90 to 4.75\,FP/hr with lower variance and a reduced inference time of 21.6\,ms at comparable parameter count; at 8,760 deployment-hours per recorder-year, the 1.57 FP/hr R90 advantage of All+SAF over All+SA corresponds to approximately 13,735 fewer high-confidence analyst reviews per recorder per year \cite{van2009management}. For applications requiring predictable false–positive budgets across sites or regulatory contexts, the MPath CBAM final–layer model (All+MPF) yields the lowest FP/hr variability (Macro~R90 CoV\,=\,0.089) at moderate cost in IR performance. Finally, for resource–constrained or embedded platforms, the All$-$D configuration provides competitive site–level accuracy with the lowest computational demand (4.87\,M parameters; 9.17\,ms inference). These results collectively indicate that no single architecture is uniformly optimal; instead, model selection should be aligned with the operational priority while maximizing discriminative power, minimizing analyst burden, ensuring stability, or meeting efficiency constraints.

\section{Conclusion}

This work introduced \textit{GetNetUPAM}, a unified benchmarking framework for evaluating performance and stability in Underwater Passive Acoustic Monitoring (UPAM). By combining nested site--year blocked cross-validation with standard random-subset evaluation, GetNetUPAM exposed deployment-relevant failure modes that conventional protocols obscured. Building on this foundation, we proposed the \textit{Adaptive Resolution Pooling and Attention Network} (ARPA--N), a lightweight architecture tailored to the irregular spectrogram dimensions of long-window UPAM data. ARPA--N integrated adaptive pooling with CBAM spatial attention to suppress global, non-biological shortcuts and emphasize localized call morphology, yielding a 14.7\% micro-AP improvement over strong 60-s baselines, with a modular architecture configurable from full-depth spatial attention for server-class monitoring to compact ablated variants for resource-constrained platforms. Together, these results showed that rigorous evaluation and efficient architectural design were mutually reinforcing: GetNetUPAM identified the environmental regimes under which attention improved representational reliability, while ARPA--N translated these insights into a detector that remained precise and operationally viable across diverse noise conditions. Thus our framework provided a reproducible foundation for research on noise-robust, deployment-ready bioacoustic detectors and supported long-term ecological monitoring and conservation objectives.

\paragraph{Future Directions} ARPA--N’s resolution-agnostic design suggested potential applicability to domains with heterogeneous time–frequency structure such as public health acoustics, environmental surveillance, and infrastructure monitoring, provided that comparable deployment-like datasets become available. Extending UPAM models to graph neural networks or other structured-attention frameworks represented an additional promising avenue, though all such architectural extensions remained untested within the scope of this study.

\section*{Acknowledgments}
\label{Ack}

This work was supported by the National Science Foundation under Grant No. \href{https://www.nsf.gov/awardsearch/showAward?AWD_ID=2346643}{\#2346643}, the U.S. Department of Defense under Award No. \href{https://dtic.dimensions.ai/details/grant/grant.14525543}{\#FA9550-23-1-0495}, and the U.S. Department of Education under Grant No. P116Z240151.
Any opinions, findings, conclusions or recommendations expressed in this material are those of the author(s) and do not necessarily reflect the views of the National Science Foundation, the U.S. Department of Defense, or the U.S. Department of Education. The author acknowledges Microsoft Copilot for assistance with grammar and structural edits, contextual extrapolation in the results section, and figure preparation with extensive human input \cite{Copilot}. All core ideas, analyses, and interpretations are the author’s own based on previous literature and experimental analysis, as cited in the manuscript.

{\footnotesize
    \bibliographystyle{IEEEtran}
    \bibliography{whales}
}

\end{document}